\documentclass[aip,reprint,amsmath,amssymb]{revtex4-1}

\usepackage{graphicx}
\usepackage{dcolumn}
\usepackage{float}
\usepackage{bm}
\usepackage{amssymb}
\usepackage{lipsum}
\usepackage{color}

\begin{document}

\title{Segregation of cylindrically confined cross-linked polymers}

\author{James M. Polson} 
\affiliation{ Department of Physics, University of Prince Edward Island, 550 University Avenue, Charlottetown, Prince Edward Island, C1A 4P3, Canada }
\author{Tyler M. Dickieson}
\affiliation{ Department of Physics, University of Prince Edward Island, 550 University Avenue, Charlottetown, Prince Edward Island, C1A 4P3, Canada }
\author{Matthew R. O'Connell}
\affiliation{ Department of Physics, University of Prince Edward Island, 550 University Avenue, Charlottetown, Prince Edward Island, C1A 4P3, Canada }

\date{\today}

\begin{abstract}
Entropic repulsion between polymer chains under confinement in a narrow channel tends to induce segregation. Recent computer simulation studies have provided indirect evidence that internal cross linking of ring polymers, which creates internal loops, enhances the entropic force. Such topologically enhanced entropic forces are thought to contribute to the segregation of bacterial chromosomes during replication. In this study, we use Monte Carlo computer simulations of a coarse-grained hard-sphere chain model to measure the configurational free energy of two cross-linked ring polymers as a function of the separation distance between the polymer centers of mass. We examine the effects of varying the polymer architecture as well as the confinement dimensions on the free energy functions. In addition, we use the results to make a quantitative assessment of the accuracy of a simple theoretical model built on the de~Gennes blob model that has been used in past studies to estimate the overlap free energy of confined polymers. Consistent with indirect evidence from previous simulations studies of segregation dynamics, we find that increasing the number of internal loops and selecting binding sites to enforce a side-loop configuration tends to increase the overlap free energy. Analysis of the whole-polymer and interior-loop extension and overlap lengths reveal a sequence of overlap domains that explain subtle variations in the free energy functions. For all architectures examined, the theoretical model provides a reasonably accurate prediction for how the cross linking and degree of confinement affects the free energy, with quantitative discrepancies comparable to those observed in previous studies using linear polymers and ring polymers without cross links.
\end{abstract}

\maketitle

\section{Introduction}
\label{sec:intro}

When two self-avoiding polymer chains overlap, there is a reduction in the number of accessible conformational states and, thus, a reduction in the conformational entropy. This results in a gradient in the free energy (the ``entropic force'') that tends to push the polymer apart.\cite{grosberg1982polymeric, dautenhahn1994monte} This effect is greatly enhanced by confining the polymers to anisometric spaces with one or more dimensions small compared to the radius of gyration of unconfined, free polymers.\cite{daoud1977statistics,jun2006entropy} Of particular note is the case of polymer confinement in narrow channels, where the entropic force drives the polymers to segregate along the axis of the channel. The segregation process has been the subject of numerous theoretical and computer simulation studies in recent years.\cite{jun2006entropy, teraoka2004computer, jun2007confined, arnold2007time, jacobsen2010demixing, jung2010overlapping, jung2012intrachain, liu2012segregation, racko2013segregation, minina2014induction, polson2014polymer, du2018polymer, polson2021free, jung2012ring, dorier2013modelling, minina2015entropic, shin2014mixing, chen2015polymer, polson2018segregation, nowicki2019segregation, nowicki2019electrostatic, mitra2022topology, mitra2022polymer, pande2024topology, bhandarkar2026entropic, doshi2026kinetics}  Most of these simulation studies employed simple coarse-grained models to characterize the segregation behavior of flexible linear polymers, \cite{jun2006entropy,teraoka2004computer,jun2007confined, arnold2007time, jacobsen2010demixing, jung2010overlapping, jung2012intrachain, liu2012segregation, racko2013segregation, minina2014induction, polson2014polymer, du2018polymer, polson2021free} as well as ring polymers. \cite{jung2012ring, dorier2013modelling, shin2014mixing, minina2015entropic, chen2015polymer, polson2018segregation} The effects of a variety of system features on the  segregation dynamics and thermodynamics have been examined, including bending rigidity,\cite{racko2013segregation, polson2014polymer} macromolecular crowding,\cite{shin2014mixing, chen2015polymer, polson2018segregation} electrostatics,\cite{nowicki2019segregation, nowicki2019electrostatic} difference in polymer contour length,\cite{polson2021free} and topological modifications.\cite{mitra2022topology, mitra2022polymer, pande2024topology, bhandarkar2026entropic, doshi2026kinetics} 

Theoretical studies of entropy-driven segregation of confined polymers are largely motivated by its relevance to the conceptually related process of chromosome segregation in self-replicating bacteria.\cite{jun2006entropy, jun2010entropy, diventura2013chromosome, youngren2014multifork, mannik2016role} In this process, the circular DNA of the chromosome splits at two replication forks that proceed bidirectionally along the contour of the circular DNA. The newly formed arms of the chromosome segregate as replication proceeds, with duplicated origins ({\it ori}) migrating to the ends of the bacterial cell while the terminus ({\it ter}) region remains at the center of the nucleoid.\cite{badrinarayanan2015bacterial, gogou2021mechanisms} 
Though some experimental studies yielded results difficult to reconcile with a purely entropic segregation mechanism,\cite{fritsche2011model, benza2012physical, yazdi2012variation, fisher2013four, diventura2013chromosome, kuwada2013mapping} some recent studies have reported results that are consistent with and offer indirect support for this hypothesis.\cite{diventura2013chromosome, mannik2016role, cass2016escherichia, wu2020geometric, elnajjar2020chromosome, japaridze2020direct} For example, Wu {\it et al.} observed that the efficiency of cell growth and chromosome segregation of a cell-wall-less variant of {\it B. subtilis} is greatly enhanced by increasing the degree of lateral confinement when constraining the cells to narrow channels.\cite{wu2020geometric} Though the effect can be rationalized by invoking an out-of-equilibrium process involving steric repulsion with polysomes whose impact is expected to weaken as the cells widen,\cite{papagiannakis2025nonequilibrium} the trend is also consistent with the observed increase in the entropic force with increasing confinement in simulation studies using simple polymer models.\cite{polson2014polymer, polson2018segregation, polson2021free}. 

In a seminal simulation study two decades ago, Jun and Mulder used a polymer model of a replicating chromosome to show that entropic demixing alone was sufficient to explain observed patterns during concurrent replication and segregation, provided the replicated arms were given a greater confinement volume than the unreplicated strand.\cite{jun2006entropy} Without employing this ``concentric shell'' model, segregation was blocked during the early stages of replication. Another study employed the constraint of fixing the replisomes at a mid-cell position and was able to reproduce the observed behavior without treating confinement in this manner.\cite{elnajjar2020chromosome} More recently, other simulation studies employing topologically modified polymers have shown that entropic forces alone can yield results consistent with experimental results without the need for either differential confinement or constraints on replisome positions.\cite{mitra2022polymer, pande2024topology, harju2024loop} The central idea is that intra-arm cross-linking creates internal loops that entropically repel one other, thus enhancing the force driving the arms apart, even at early stages of replication. 

Of particular interest to the present study is the work carried out by Chatterji and coworkers,\cite{mitra2022polymer, mitra2022topology, pande2024topology, bhandarkar2026entropic, doshi2026kinetics} who have conducted out a series of related simulation studies investigating the entropic effects of internal loops created by cross links at {\it fixed} locations. Employing architectures judiciously chosen to reproduce both the linear ordering of genomic loci and the patterns of the Hi-C contact maps observed in {\it in vivo} experiments of {\it E. Coli}, they found an enhanced propensity for segregation under both slow-growth conditions, where there are only two replications forks present at any given time,\cite{mitra2022polymer} and under fast-growth conditions, where additional replication forks are present.\cite{pande2024topology} In addition, these researchers have examined the spatial organization as well as the  segregation propensity and kinetics of initially overlapping, fully replicated polymers with a wide range of internally-looped architectures.\cite{mitra2022topology, doshi2026kinetics} In Ref.~\onlinecite{mitra2022topology}, they found that the details of the architecture profoundly affected the segregation success rate. Furthermore, a simple extension of the de~Gennes blob model was used to estimate the entropic force strength for each architecture, yielding results that correlate with the propensity for segregation. In  Ref.~\onlinecite{doshi2026kinetics}, they examined the kinetics of segregation for a collection of related polymer architectures characterized by a variable number of loops of fixed total length on one side of the polymer ring.\cite{doshi2026kinetics} They found that increasing the number of loops (and simultaneously decreasing their size) tended to increase the segregation rate.

The main molecular architectures employed in Refs.~\onlinecite{mitra2022polymer, pande2024topology} were designed in accord with the experimental observation of aggregation of {\it rrn} operons located at genomically distant chromosomal loci in {\it E. coli}.\cite{gaal2016colocalization} This aggregation may be driven by clustering of RNA polymerase (RNAP) through the mechanism of liquid-liquid phase separation.\cite{ladouceur2020clusters}
Such a clustering of {\it rrn} operons could function as an effectively fixed cross link over genomic distances much greater than those for expected for cross links arising from nucleoid associated proteins. The model architecture labeled Arc-2 have cross links located positions that roughly correspond to the positions of specific operons known to be colocalized {\it in vivo}.\cite{gaal2016colocalization} A modified version of this architecture, called Arc-2-2, has additional cross links that were introduced to obtain a contact map similar to the Hi-C map for {\it E. coli} while also localizing genomic loci in accord with observations in Flourescence {\it in situ} hybridization (FISH) experiments.

Effective cross links between genomically distant parts of bacterial chromosomes also arise from the action of Structural Maintenance of Chromosomes (SMC) complexes. SMCs are motor proteins that attach to the chromosome and then progressively reel in a chromosomal loop in an ATP-driven process. However, in this case the cross links move along the chain contour sufficiently fast that they cannot be modelled as a fixed links on the timescale of chromosome segregation. Experimental studies have demonstrated that SMC complexes are required for faithful chromosome segregation in certain bacteria such as {\it E. coli},\cite{burmann2025mechanism} which lack the parABS system that drives origin separation. Simulations employing data-driven models support this finding.\cite{brahmachari2026exploring} Likewise, SMCs have been shown to contribute significantly to segregation in other bacteria.\cite{maytin2026chromosome, pawlikiewicz2026smc, izquierdo2025chromosome} A recent simulation study by Harju {\it et al.}\cite{harju2024loop} investigated the effect of SMCs on bacterial chromosome segregation. They considered the case of {\it dynamical} cross links that mimic the effect of SMCs loaded at the {\it ori} positions of newly formed chromosomal arms of a replicating model chromosome, as is the case for bacteria such as {\it B. subtilis} and {\it C. crescentus}. The cross links, moving along the arms away from the {\it ori} sites, were found to enhance the entropic driving force, especially so at early stages of replication, where entropic forces were found to {\it inhibit} segregation in the absence of cross linking. 

In this study, we examine the effects of topological modifications on the entropic forces driving segregation of coarse-grained model polymers under cylindrical confinement. In particular, we consider the case of polymers with cross links located at {\it fixed} locations along the chains and consider a few molecular architectures that were the focus of studies of Refs.~\onlinecite{mitra2022topology} and \onlinecite{doshi2026kinetics}. 
The architectures and the cross-link sites of the polymers are illustrated in Fig.~\ref{fig:archpol_illust}.
As in those works, we consider segregation of two fully replicated polymers, rather than examining concurrent replication and segregation as in Refs.~\onlinecite{mitra2022polymer, pande2024topology, harju2024loop}, a case that will be considered in future work. We employ Monte Carlo (MC) simulations to measure the variation of the configurational free energy with respect to the distance between the centers of mass of the chains. This work builds on previous studies that characterized such free-energy functions for cylindrically confined linear and ring polymers.\cite{polson2014polymer, polson2018segregation, polson2021free} We examine the effects on the free energy of variations in the molecular architecture as well as the confinement dimensions of both infinite-length and finite-length cylinders. As the gradient in the free energy is a measure of the entropic force, these functions can provide a relative measure of the propensity for segregation as well as the segregation rates. As in our previous work, the overlap free-energy barrier scales with channel diameter in a manner roughly consistent with the predictions using a modified form of the blob model, with small quantitative discrepancies comparable in magnitude to those seen for linear polymers and ring polymers without cross links. In addition, the free energy barrier increases with increasing topological complexity, in semi-quantitative agreement with the blob model and  in qualitative agreement with the results for segregation success rates for the architectures examined in Ref.~\onlinecite{mitra2022topology}. 

The remainder of the article is organized as follows. Section~\ref{sec:model} describes the model employed in the simulations. Section~\ref{sec:methods} describes in detail the MC procedure used to carry out the simulations. Section~\ref{sec:results} presents the results of the calculations. Finally, Sec.~\ref{sec:conclusions} summarizes the main conclusions of this work.

\section{Model} 
\label{sec:model}

The system consists of two identical ring polymers confined to a cylindrical channel or cavity. Each  polymer is modelled as a freely-jointed chain of $N=500$  hard spheres of diameter $\sigma$. The pair potential for each pair of nonbonded particles is thus $u_{\rm nb}=\infty$ for $r\leq\sigma$, and $u_{\rm nb}=0$ for $r>\sigma$, where $r$ is the distance between the centers of the monomers. Pairs of bonded monomers interact with a potential of $u_{\rm b} = 0$ if $0.9\sigma < r < 1.1\sigma$, and $u_{\rm b} = \infty$ otherwise. In addition to standard bonding between monomers along the chain, selected pairs of sites are also subject to a cross-linking potential, given by $u_{\rm cl}=0$ if $\sigma < r < 1.5\sigma$, and $u_{\rm cl}=\infty$, otherwise. We choose the range for finite $u_{\rm cl}$ to be somewhat larger than that for $u_{\rm b}$ (i.e. $0.5\sigma$ vs $0.2\sigma$) to compensate for an observed reduction in computational efficiency for the MC crankshaft trial moves at the cross-linked sites.

\begin{figure}[!ht]
\begin{center}
\vspace*{0.0in}
\includegraphics[width=0.45\textwidth]{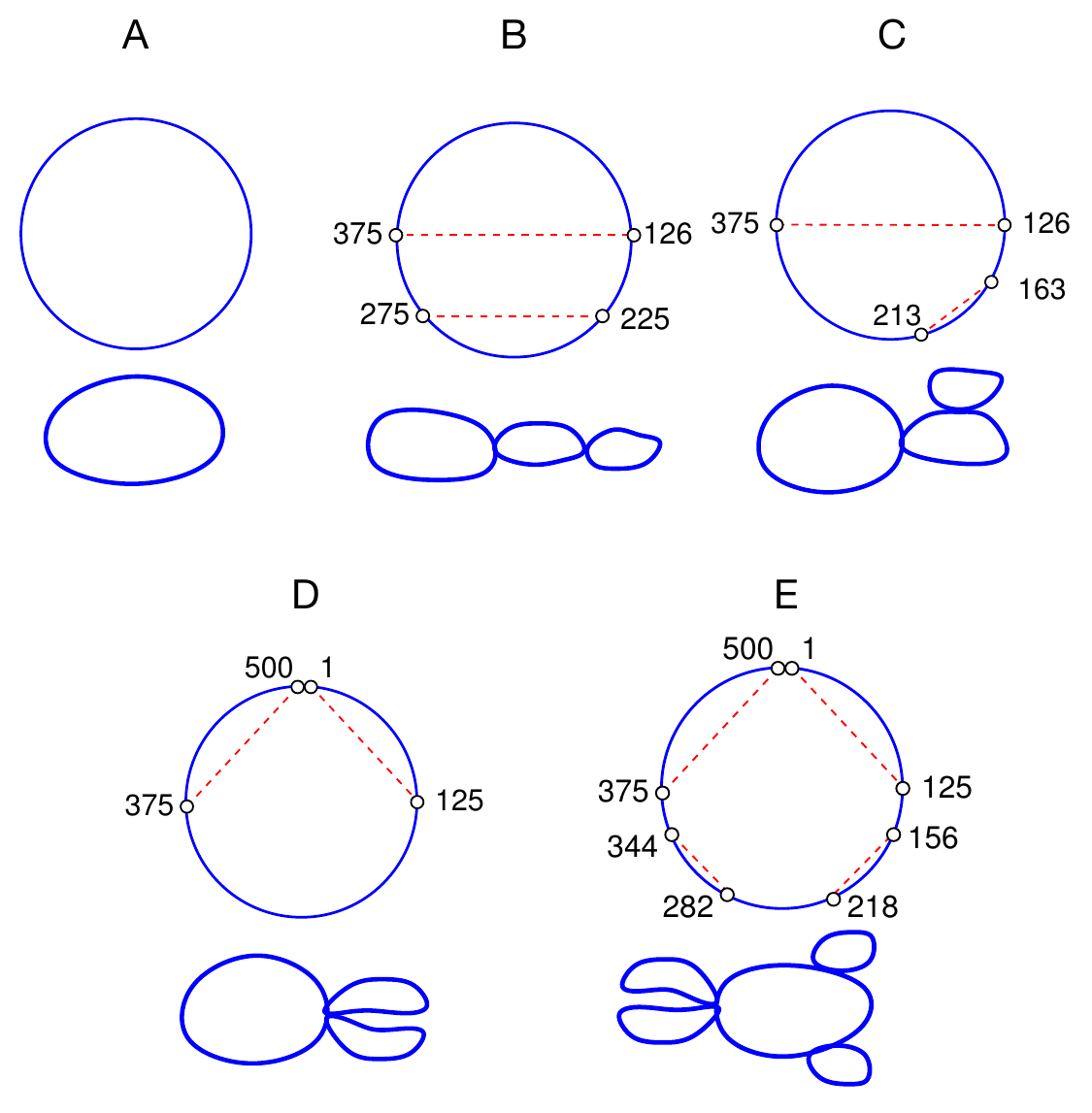}
\vspace*{-0.2in}
\end{center}
\caption{Illustration of the five polymer architectures examined in this study. The top portion of each part of the figure shows the cross linking sites of the ring polymer, and the bottom portion illustrates the resulting topology.  Dashed lines between pairs of labelled monomer sites on the chain denote cross linking between those monomers.}
\label{fig:archpol_illust}
\end{figure}

Five different polymer architectures were considered in this study, some of which have been examined previously by Chatterji and coworkers.\cite{mitra2022topology, mitra2022polymer, pande2024topology, bhandarkar2026entropic, doshi2026kinetics} These specific architectures were selected to highlight the effects of varying the number and placement of the internal loops on the free-energy functions. Generally, this entropic repulsion tends to increase with increasing the number of internal loops and placing them in positions that force intra-polymer overlap between loops. Each of these architectures are defined by the pairs of sites on the ring polymer that are cross linked. Figure~\ref{fig:archpol_illust} illustrates the architectures, showing the identities of the monomers that are cross linked as well as the resulting topology that arises from the cross linking. We use a nomenclature similar to but somewhat simpler than those used in Refs.~\onlinecite{mitra2022topology, mitra2022polymer, pande2024topology, bhandarkar2026entropic, doshi2026kinetics}. Arch-A is a simple ring polymer with no cross linking. (Note that ``Arch'' denotes ``architecture''.)
It was labelled Arc-0 in Refs.~\onlinecite{mitra2022topology, doshi2026kinetics}. Arch-B has two cross links chosen with a symmetry that results in three loops that tend to organize linearly when confined to a cylinder. It has no exact equivalent in these previous studies. In contrast to Arch-B, the asymmetrical positions of the two cross links in Arch-C result in a side loop that tends to overlap with another loop on the same polymer under cylindrical confinement. While Arch-C is topologically equivalent to the ``linearly looped" case of Arc-11 in Ref.~\onlinecite{mitra2022topology}, the binding sites defining the placement of the side loop forces it to overlap fully with the middle loop along the channel. Consequently, Arch-C better resembles the ``side-looped" case of Arc-4.  Arch-D also has two pairs of cross-linked monomers with three internal loops; however, the choice of linking sites  results in two identical small loops, each connected to a larger loop at essentially the same location. Arch-D is the one architecture to appear in each of the studies of Refs.~\onlinecite{mitra2022topology, mitra2022polymer, pande2024topology, bhandarkar2026entropic, doshi2026kinetics}, where it was named Arc-2,\cite{mitra2022topology, mitra2022polymer, pande2024topology} or Arc1-2.\cite{bhandarkar2026entropic, doshi2026kinetics} Finally, Arch-E is a modification of Arch-D with two additional cross links that result in two additional internal loops that have a tendency to overlap (partially) with the large loop under cylindrical confinement. Arch-E is essentially the same as Arc2-2 in Refs.~\onlinecite{mitra2022topology, mitra2022polymer}.  As noted in Sec.~\ref{sec:intro}, these latter two architectures were designed to reproduce the observed linear ordering of genomic loci and the patterns of Hi-C contact maps observed in {\it in vivo} experiments of {\it E. coli}. Most of the simulations in this work focus on Arch-D. Some simulations examine the effects of changing the locations of the binding sites of Arch-D in a manner that preserves the topology and symmetry but changes the relative contour lengths of the large and small loops. Note that in all simulations, the loops of the polymers are topologically constrained to be uncatenated. Thus, unlike the simulations in e.g. Ref.~\onlinecite{mitra2022topology}, we do not consider effects of strand passage effected in bacterial chromosomes by topoisomerase. Also note that the two polymers are disconnected, in contrast to the case in Ref.~\onlinecite{mitra2022topology}, where the polymers were linked at sites corresponding to the {\it ter} positions in bacterial chromosomes. 

The polymers are confined to a hard-wall cylindrical channel that is aligned along the $z$-axis. 
The channel has an effective diameter $D$ such that each monomer interacts with the wall of 
the tube with a potential $u_{\rm w}(r) = 0$ for $r<D$ and $u_{\rm w}(r) = \infty$ for $r>D$, 
where $r$ is the distance of the monomer center from the central axis of the cylinder. Thus, 
$D$ is the diameter of the cylindrical volume accessible to the centers of the monomers, and
the actual diameter of the cylinder is $D+\sigma$. Some simulations employed cylindrical cavities of finite length. In these cases, the cylinders are capped with hemispheres with a radius equal to that of the cylinder. The cavity length $L$ is defined as the length of cylinder portion of the spherocylinder. 

\section{Methods}
\label{sec:methods}

In most simulations we measure the variation of the free energy with respect to $\lambda$, the distance between the polymer centers of mass along the cylinder axis.  We employ Monte Carlo simulations with the Metropolis algorithm together with the self-consistent histogram (SCH) method,\cite{frenkel2002understanding} a multiple-histogram method we have used in previous studies of polymer segregation.\cite{polson2014polymer, polson2018segregation, polson2021free}  (Note that the SCH method is closely related to the Weighted Histogram Approximation Method (WHAM).\cite{kumar1992weighted}) To implement the SCH method, we carry out many independent simulations, each of which employs a unique ``window potential'' of the form:
\begin{eqnarray}
{W_i(\lambda)}=\begin{cases} \infty, \hspace{8mm} \lambda<\lambda_i^{\rm min} \cr 0,
\hspace{1cm} \lambda_i^{\rm min}<\lambda<\lambda_i^{\rm max} \cr \infty,
\hspace{8mm} \lambda>\lambda_i^{\rm max} \cr
\end{cases}
\label{eq:winPot}
\end{eqnarray}
where $\lambda_i^{\rm min}$ and $\lambda_i^{\rm max}$ are the limits that define the range of $\lambda$ for the $i$-th window.  Within each window of $\lambda$, a probability distribution $p_i(\lambda)$ is calculated in the simulation. The window potential width, $\Delta \lambda \equiv \lambda_i^{\rm max} - \lambda_i^{\rm min}$, is chosen to be sufficiently small that the variation in $F$ does not exceed a few $k_{\rm B}T$ within the window bounds. The windows are chosen to overlap with half of the adjacent window, such that $\lambda^{\rm max}_{i} = \lambda^{\rm min}_{i+2}$.  The window width was typically in the range $\Delta \lambda = 2\sigma-4\sigma$. The SCH algorithm was employed to reconstruct the unbiased distribution, ${\cal P}(\lambda)$, from the $p_i(\lambda)$ histograms. The free energy follows from the relation $F(\lambda) = -k_{\rm B}T\ln {\cal P}(\lambda)+{\rm const}$. The constant is chosen for convenience and for visual clarity of the graphs. For the case where the cylinder length is infinite, it is usually convenient to choose $F(\lambda=\infty)=0$, whereas for finite-length channels, it is convenient to choose $F(\lambda=\lambda_{\rm max})=0$, where $\lambda_{\rm max}$ is the separation distance where $F$ is a maximum.

Polymer configurations were generated by carrying out single-monomer moves consisting of either a translational displacement or a crankshaft rotation. In addition, whole-polymer-displacement moves of each polymer along the channel axis were also employed to increase the efficiency of sampling $p_i(\lambda)$.  Each trial move was rejected if it resulted in overlap between particles or between a particle and a confinement surface, or if it led to a violation of the constraints associated with bonding, cross-linking or the SCH window potentials; otherwise it was accepted. 

Each simulation began with a multi-step initialization procedure employing the Metropolis MC algorithm.  At the beginning of the simulation, the two ring polymers, each $N$ monomers in length, were placed in highly extended conformations that satisfied the bonding constraints as well as the condition of non-overlap between pairs of non-bonded monomers and between monomers and the surface of the confining cylinder. The window potential of Eq.~(\ref{eq:winPot}) was initially turned off, and the centers of mass of the two chains were  initially coincident along the channel (i.e., $\lambda=0$), and so the constraint on $\lambda$ imposed by the window potential was typically not initially satisfied. In addition, the cross-linking constraints were also not initially in place. To create a cross link between a pair of monomers, an attractive pair potential between them was imposed in order to bring those monomers together. In order to prevent catenation between loops on different polymers, the two polymers were held laterally separated by imposing a constraint that the monomers of each polymer remain in a separate longitudinal half of the cylinder (i.e., monomers for one polymer were constrained to be at locations with $y>0$, and those of the other polymer were forced to be at positions with $y<0$.). A comparable approach was used to prevent intra-polymer loop catenation. Once the monomers were separated by a distance within the range of the binding potential, the attractive force for that pair was turned off, and the binding potential was turned on. This phase of initialization ended when all the cross links were in place, following which the constraint for lateral separation was turned off. 

In the case of finite channel length, the second phase of initialization involved a reduction in the length of the cylinder, which was initially set to a value greater than the extension length of the polymer in its initial configuration. At regular intervals, an attempt to reduce the length of cylinder was made. If the trial value of the cylinder length did not lead to overlap with any monomers, the new value was accepted. This phase of initialization ended when the cylinder length reached its target value. The final stage of initialization involved moving the polymers into configurations that satisfied the separation constraint imposed by the window potential of Eq.~(\ref{eq:winPot}). To do this, a ``guiding'' potential energy dependent on $\lambda$ was implemented. This energy decreased when $\lambda$ came closer to the target range and increased as it went further away. Eventually, $\lambda$ satisfied the separation constraint, following which the guiding potential was turned off and the window potential was turned on. At this point, all of the required constraints on the polymer configurations were in place.

Following initialization, the system was equilibrated for $\mathcal{O}(10^{7})$ MC cycles, following which various quantities of interest were sampled every 100 MC cycles over a run time of ${\cal O}(10^8)$ MC cycles. Each MC cycle corresponds to $N$ randomly selected trial moves corresponding to either a single-monomer translation or crankshaft rotation or a whole-polymer displacement. On average, $49\%$ of the trial moves were monomer displacement moves, $49\%$ were crankshaft moves, and $2\%$ were whole polymer moves. The following maximum displacements for trial moves were chosen: $\Delta_{\textrm{max}}=0.22\sigma$ for single-monomer displacements, $\theta_{\textrm{max}}=\pi$ for crankshaft rotations, and $Z_{\textrm{max}}=0.1\sigma$ for whole-polymer translational moves. These values yielded acceptance ratios in the range $30-50\%$. Note that the maximum displacements of the single-monomer and whole-polymer moves were chosen to be sufficiently small to prevent any chain-crossing events. Thus, the internal loops of the polymers were topologically constrained to be uncatenated. 

In all simulations, the number of monomers per polymer was chosen to be $N=500$. The diameter of the cylindrical channel was varied, with values in the range $D=11\sigma-17\sigma$. In most of the simulations, the channel length, $L$, was infinite. In cases of finite-length cylinders, the channel length was in the range of $L=60\sigma-80\sigma$.  The particular monomers that were cross linked
were selected for each polymer architecture in accord with Fig.~\ref{fig:archpol_illust}.

Each window is defined by a unique choice of values of $\lambda_i^{\rm min}$ and $\lambda_i^{\rm max}$ for the potential of Eq.~\eqref{eq:winPot}, and the number of windows depends on the desired range of $\lambda$ for the free energy function. For example, simulations for channels of length $L=80\sigma$ typically used 29 windows. Multiple statistically independent simulations using different random-number sequences were run for each window. The probability histograms from each of these simulations were then averaged to obtain better statistics. Typically, 10-20 individual simulations were run for each window. Thus, to obtain a single free energy function, a set of 300-500 simulations were run, each of which required 1.5 to 2.5 days of CPU time.

In the results presented below, distances are measured in units of $\sigma$, and energy is measured in units of $k_{\rm B}T$.

\section{Results}
\label{sec:results}

\subsection{Infinite-length channels}
\label{subsec:infin}

In this section we characterize the physical state of the polymers upon variation of the center-of-mass displacement $\lambda$ for confinement in infinitely-long channels. While the central quantity is the free energy, we also examine the variation of extension lengths and overlap distances for the entire polymers and for individual internal loops for additional insights. Most of the results presented here pertain to Arch-D chains, the architecture most examined in several recent studies.\cite{mitra2022topology, mitra2022polymer, pande2024topology, bhandarkar2026entropic, doshi2026kinetics} In order to help understand the effects of cross-linking in these chains, it is helpful to first consider the simpler case of Arch-A chains, i.e. ring polymers with no cross linking. 

\begin{figure}[!ht]
\begin{center}
\vspace*{0.0in}
\includegraphics[width=0.48\textwidth]{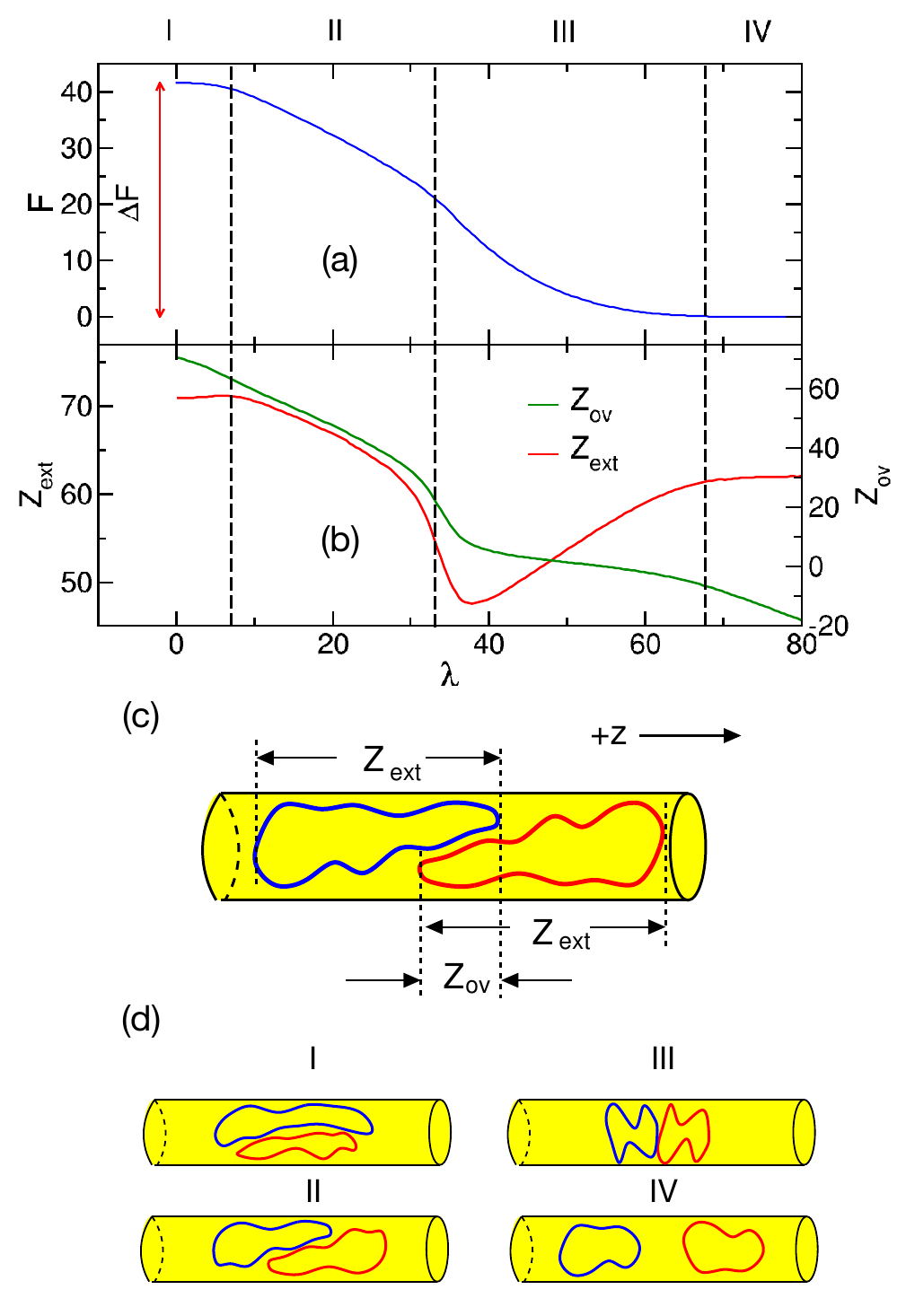}
\vspace*{-0.2in}
\end{center}
\caption{Illustration of the overlap regimes for two polymers of type Arch-A (i.e. simple ring polymers) confined to an infinite channel of diameter $D=15$. (a) Free energy vs $\lambda$. (b) Extension length, $Z_{\rm ext}$, and  overlap length, $Z_{\rm ov}$, vs $\lambda$. (c) Visual illustration of the definitions of  $Z_{\rm ext}$, and  $Z_{\rm ov}$. (d) Illustration of polymer configurations for each overlap regime.}
\label{fig:archA_regimes}
\end{figure}

Figure~\ref{fig:archA_regimes}(a) shows the variation of the free energy with respect to the center-of-mass separation distance $\lambda$ for two such ring polymers. Figure~\ref{fig:archA_regimes}(b) shows the corresponding variation of the polymer extension length and the overlap distance for the same system. As noted in previous studies,\cite{polson2014polymer, polson2018segregation, polson2021free} there are four physically distinct regimes. Regime~I lies at very low values of $\lambda$, the free energy and the average extension length are roughly constant. In this regime, one polymer is slightly compressed relative to the other and completely nested within it. A small free energy barrier separates this state from another state where the two polymers trade places (i.e. the extended polymer switches to the nested polymer and vice versa). The effect has been discussed at length in Refs.~\onlinecite{polson2018segregation} and \onlinecite{polson2021free}. At somewhat higher $\lambda$, the system enters regime~II, which is characterized by partial overlap of the polymers along the channel.  Here, the free energy decreases monotonically with $\lambda$, as do the extension length per polymer and the overlap length. The extension length decreases because the non-overlapping portion of the polymer has more lateral space to occupy compared to the overlapping portion, and consequently is less extended along the channel. As $\lambda$ increases further, the  extension length and the overlap length both abruptly decrease, and the system enters regime~III. This is characterized by a near constant overlap distance of approximately zero and an extension length that increases with $\lambda$. There is also a visible change in the curvature of $F(\lambda)$, though $F$ still decreases monotonically with $\lambda$. In this regime, the polymers are essentially segregated but compressed and in contact, analogous to two compressed springs whose extension lengths increase as the distance between their centers increases. Finally, at sufficiently high $\lambda$, the system enters regime~IV, characterized by segregated polymers that are not touching. Here, further increasing $\lambda$ cannot affect the conformational entropy or the extension length, which therefore remain constant. As discussed in our earlier work,\cite{polson2014polymer, polson2018segregation, polson2021free} the variation of the free energy and extension and overlap lengths with respect to $\lambda$ can be semi-quantitatively accounted for using a theoretical model that employs the de~Gennes blob model\cite{deGennes_book} together with a simple approximation for the  free energy of the portions of the polymers that overlap in the channels.\cite{jung2012ring} 

\begin{figure*}[!ht]
\begin{center}
\includegraphics[width=0.95\linewidth]{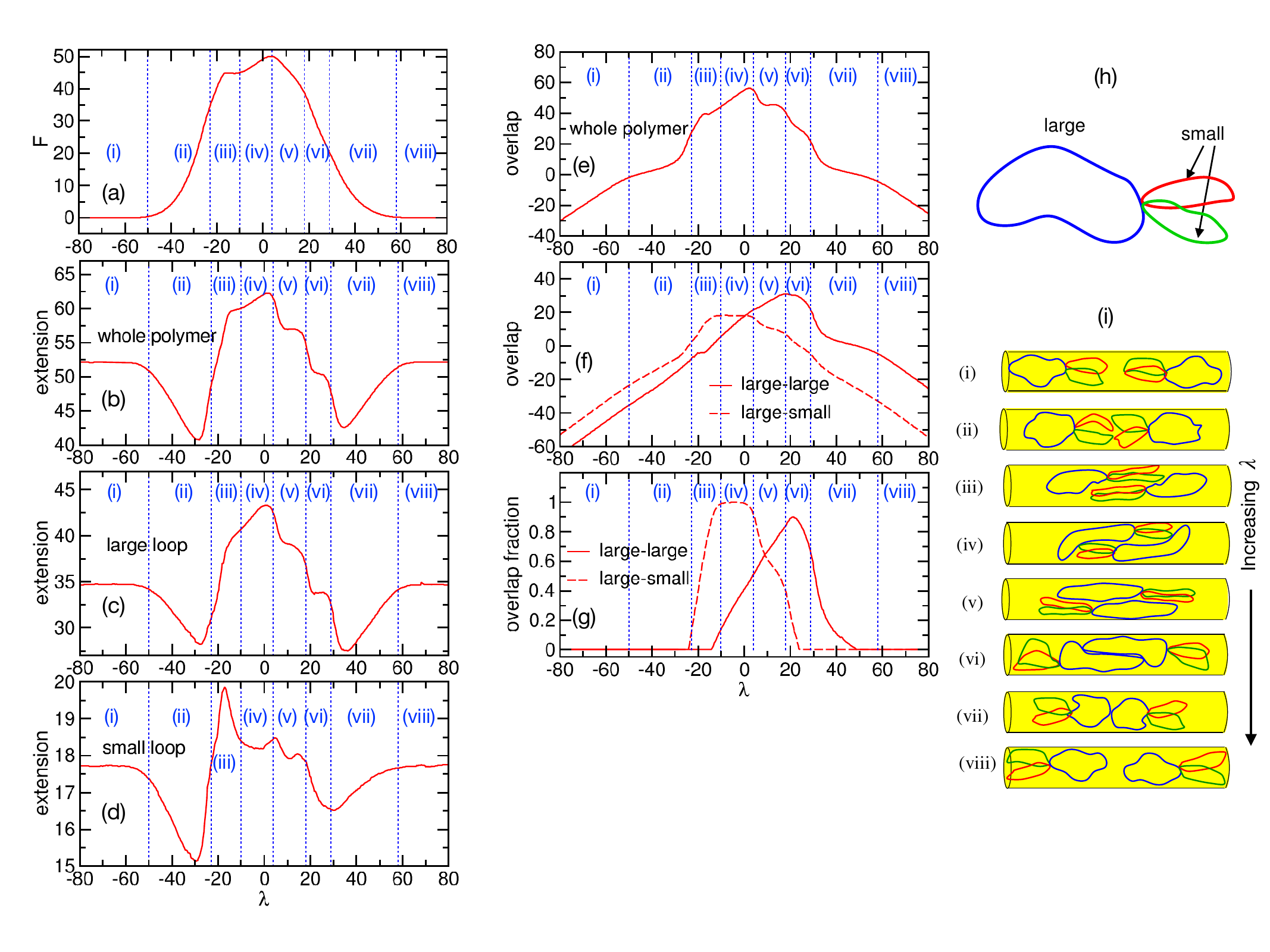}
\vspace*{-0.3in}
\end{center}
\caption{Free energy, extension lengths, and overlap lengths and fractions vs $\lambda$ for two antiparallel Arch-D chains in a channel of diameter $D=13$. (a) Free energy function.  (b) Extension length of each polymer. (c) Extension length of the large loop in each polymer. (d) Extension length of the small loops in each polymer. (e) Polymer-polymer overlap length.  (f) Inter-polymer overlap length of the large loops and of the large loops with small loops. (g) Fraction of the large loop and of the small loops on one polymer that overlaps with the large loop of the other polymer. (h) Loop labeling for the Arch-D polymer. (i) Illustration of the overlap regimes for the polymers. The inter-polymer displacement, $\lambda$ increases from regime (i) through to regime (vii).}
\label{fig:D_mirror_ext_lov}
\end{figure*}

\begin{figure*}[!ht]
\begin{center}
\includegraphics[width=0.95\linewidth]{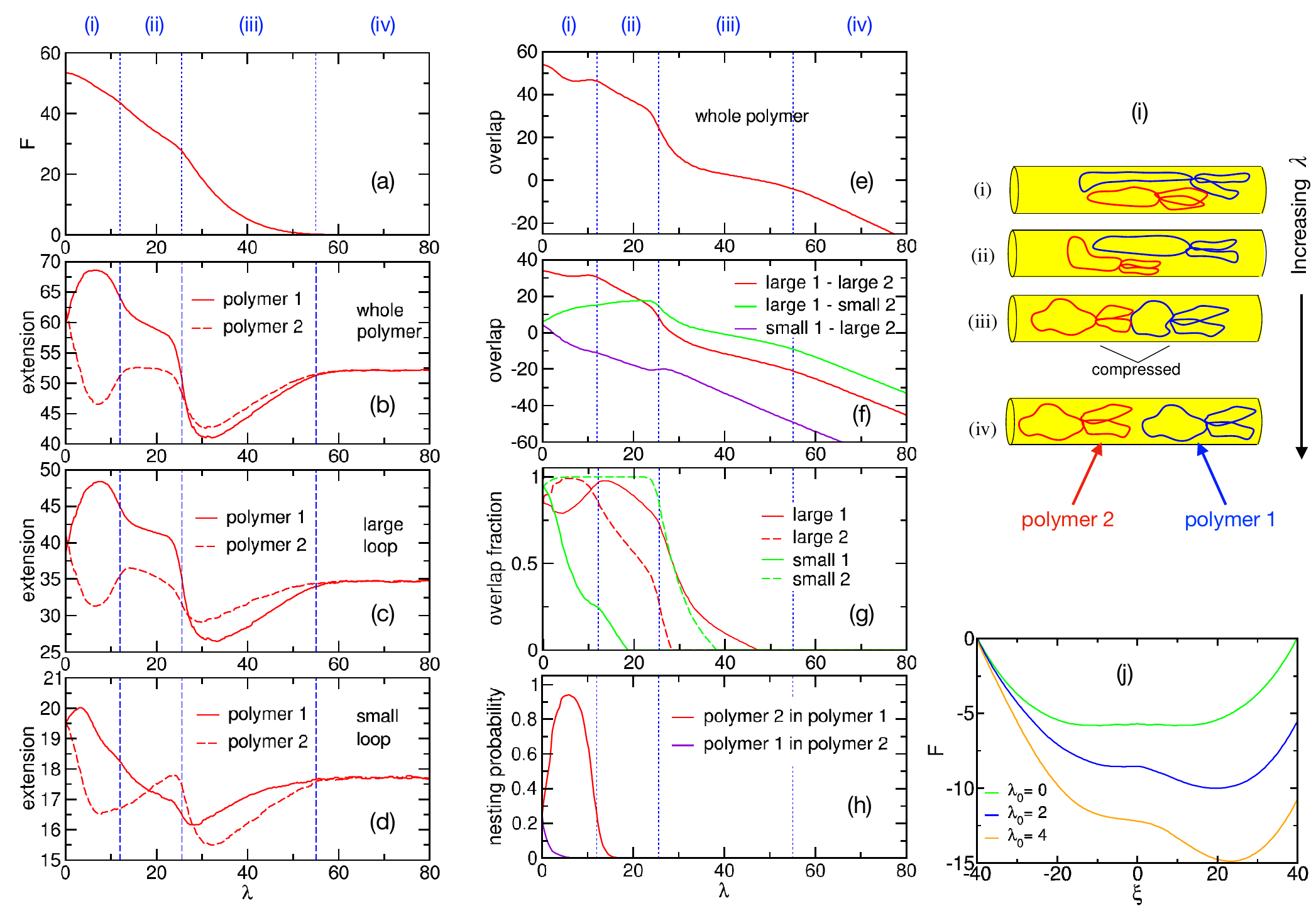}
\vspace*{-0.3in}
\end{center}
\caption{Free energy, extension lengths and overlap length vs $\lambda$ for parallel Arch-D chains in a channel of diameter $D=13$. (a) Free energy. (b) Extension length of each polymer. (c) Extension length of the large loop in each polymer. (d) Extension length of the small loops in each polymer. (e) Polymer-polymer overlap length.  (f) Inter-polymer overlap length for the different loops. (g) Fraction of the large and small loops on one polymer that overlaps with any portion of the other polymer. (h) Probability that one polymer is nested inside the other. (i) Illustration of the overlap regimes for the polymers.  (j) Free energy vs difference in polymer extension lengths, $\zeta\equiv Z_{{\rm ext,}1}-Z_{{\rm ext},2}$ for cases where the polymer centre-of-mass displacement constrained to lie in $[\lambda_0-2,\lambda_0+2]$. Results are shown for several $\lambda_0$. }
\label{fig:D_nonmirror_ext_lov}
\end{figure*}

Unlike the simple ring polymers of Arch-A, the cross linking in an Arch-D chain creates a topological asymmetry that leads to distinct orientations when it is confined to a cylinder.\cite{bhandarkar2026entropic} The polymer tends to extend along the channel with the large loop in one region and the two smaller loops co-localized in another region. The polymer can flip between two states corresponding to two different large/small (L/S) orientations by crossing a free-energy barrier.\cite{bhandarkar2026entropic} Two overlapping Arch-D polymers also display the same tendency for intra-polymer separation of the large loops from the small loops. This creates two fundamentally different configurations: one with a {\it antiparallel} alignment of the L/S domains, and one with {\it parallel} alignment. These are illustrated in Figs.~\ref{fig:D_mirror_ext_lov}(i) and \ref{fig:D_nonmirror_ext_lov}(i), respectively. Provided that the confinement is sufficiently strong, the barriers will be high enough to make alignment flipping a rare event. In that case, the parallel and antiparallel configurations are essentially locked in place and will be characterized by distinct overlap free energy functions. However, interconversion between the two configurations complicates the calculation. If the flipping rate is slow but still non-negligible, the computational efficiency will suffer by the need to run the simulations much longer to allow a large number of flipping events. Even if this computationally feasible, the validity of using just a single reaction coordinate becomes dubious, and a more physically relevant approach would involve using at least two reactions coordinates, for example one that distinguishes between orientational configurations in addition to $\lambda$. However, while such a calculation is possible using e.g. the WHAM2D method,\cite{kumar1995multidimensional} it would likely involve a prohibitively expensive simulation. 

To investigate this behavior, we measured the free-energy barrier height for orientation flipping of a single Arch-D chain and of two Arch-D chains constrained to have near-overlapping centers of mass (i.e. $\lambda\approx 0$). The calculations are described in detail in Appendix~\ref{app:a}. For both cases, the free-energy barrier height grows with decreasing $D$, as seen in Fig.~\ref{fig:F_flip}. Interestingly, in the case of two overlapping chains, the free energy of the antiparallel state is lower than that of the parallel configuration. The smallest free-energy barrier height in the two-chain system is that for transitioning over the barrier from the parallel state. A ``high'' barrier could be defined as one where the likelihood of flipping becomes very unlikely during the process of segregation starting from $\lambda=0$. In this case, the free-energy function $F(\lambda)$ calculated with an imposed constraint that fixes the relative orientation of the chains would be most relevant to the kinetic pathway followed by the system during segregation. For the narrowest channels considered ($D=11$), the smallest free energy barrier height (i.e. the one associated with the minimum $F$ for the parallel configuration) is very likely sufficiently high to satisfy this condition. However, for wider channels, lower barrier heights make it increasingly likely that flipping will occur, in which case the system would transition to a different kinetic pathway characterized by a different entropic driving force. 

\begin{figure}[!ht]
\begin{center}
\vspace*{0.0in}
\includegraphics[width=0.48\textwidth]{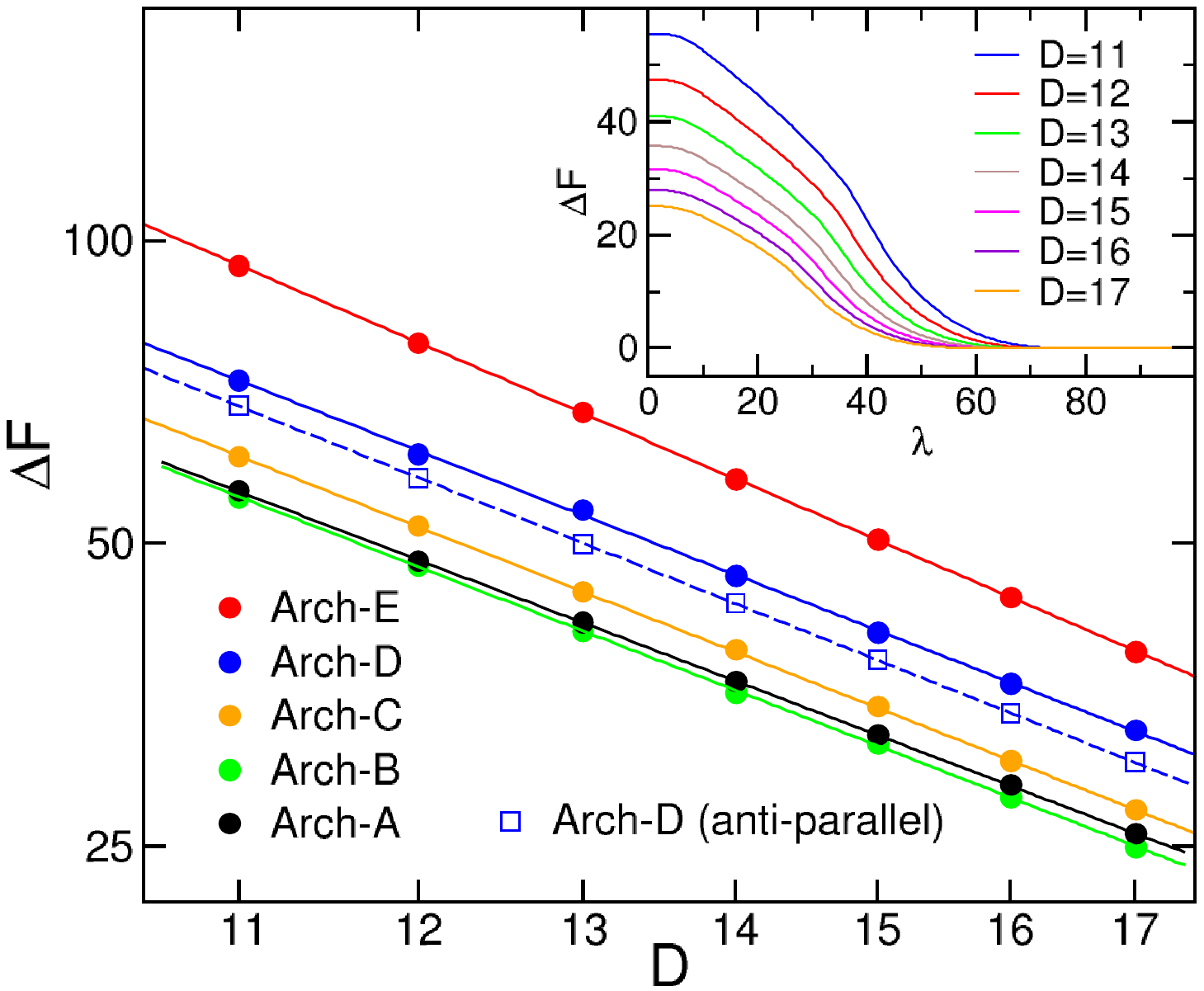}
\vspace*{-0.2in}
\end{center}
\caption{Free energy barrier height, $\Delta F$, vs channel diameter, $D$, for polymers in the parallel configuration (filled circles) confined to infinite-length channels. Results are shown for each polymer architecture. The lines show fits to a power-law function, $\Delta F \propto D^{-\beta}$. The fits yield scaling exponents of $\beta=1.80\pm 0.01$ for Arch-A, $\beta=1.84\pm 0.01$ for Arch-B, $\beta=1.86\pm 0.01$ for Arch-C, $\beta=1.84\pm 0.02$ for Arch-D, and $\beta=2.03\pm 0.01$ for Arch-E. Overlaid on these data are results for Arch-D in the antiparallel configuration (open squares), the fit to which yields a scaling exponent of $\beta=1.87\pm 0.01$. The inset shows the variation of $F$ with $\lambda$ for polymer chains of type Arch-B for several values of the channel diameter, $D$. }
\label{fig:delFvsD}
\end{figure}

For the results presented below, we carry out simulations with a constraint imposed on each polymer that the center of mass never crosses the site where the small loops are attached (analogous for constraining $\xi_1>0$, where $\xi_1$ is defined and illustrated in Fig.~\ref{fig:F_flip} in Appendix~\ref{app:a}). Free-energy functions for parallel and anti-parallel configurations are then calculated separately. We focus on a channel width of $D=13$, which likely narrow enough to satisfy the condition described above that flipping is unimportant during segregation most of the time, in which case the strength of the resulting entropic force, $|dF/d\lambda|$, is expected to correlate with segregation rate, as observed in our previous studies of linear polymers.\cite{polson2014polymer, polson2021free} In some calculations (i.e. those shown in Fig.~\ref{fig:delFvsD}) we employ cylinder diameters for which this condition is only marginally satsified. However, in almost all cases, the effect of imposing the constraint tends to have negligible effect on the results.

Figure~\ref{fig:D_mirror_ext_lov} shows a collection of results for the case of two Arch-D chains confined to an infinite cylinder of diameter $D=13$ in the antiparallel configuration. Figure~\ref{fig:D_mirror_ext_lov}(a) shows the free energy function. Figures~\ref{fig:D_mirror_ext_lov}(b), (c) and (d) show the extension lengths of the whole polymer, the large loop and the small loops, respectively. Figure~\ref{fig:D_mirror_ext_lov}(e) shows the overlap $Z_{\rm ov}$ between the whole polymers, while Fig.~\ref{fig:D_mirror_ext_lov}(f) shows the overlap between the two large loops and between the large and small loops of the different polymers. Note that the inter-polymer separation displacement, $\lambda$, is defined so that at large negative $\lambda$ the two large loops on the two polymers are further apart than the pairs of small loops are, and vice versa for large positive $\lambda$. This is evident in the illustration in Fig.~\ref{fig:D_mirror_ext_lov}(i). There are several notable features. Most fundamental is the fact that all curves are asymmetric about $\lambda=0$, a result of the combination of the asymmetry of each of these cross-linked polymers and their opposing orientation directions in the channel. On the other hand, such configurations also result in identical values of extension length, overlap distance, etc., for each polymer. Also notable are the two dips in the longitudinal extension of the whole polymer (Fig.~\ref{fig:D_mirror_ext_lov}(e)) and individual loops (Figs.~\ref{fig:D_mirror_ext_lov}(c) and (d)) at intermediate separation distances in $\lambda>0$  and $\lambda<0$ regions. These regimes are reminiscent of regime~III in Fig.~\ref{fig:archA_regimes}, and, indeed, the configurational state of the cross-linked polymer is here likewise one of simultaneously segregated and longitudinally compressed chains. Other notable features are the various shoulders in the extension lengths for the whole polymer (Fig.~\ref{fig:D_mirror_ext_lov}(b))and for the large loops ((Fig.~\ref{fig:D_mirror_ext_lov}(c)), as well as the overlap of the whole polymer (Fig.~\ref{fig:D_mirror_ext_lov}(e)), features that are notably absent in Fig.~\ref{fig:archA_regimes}. 

As in the case of polymers without internal cross linking, it is useful to divide the system into different overlap/separation regimes to better understand its configurational behavior. Here, we identify eight different regimes, each of which is labelled in Figs.~\ref{fig:D_mirror_ext_lov}(a)--(g). The locations of the boundaries between these regimes are somewhat ``fuzzy'', with specific values of $\lambda$ selected by visual inspection based on the locations of the various features in the graphs that signal  changes in overlap states. The regimes are defined by the behavior of the various extension lengths in Figs.~\ref{fig:D_mirror_ext_lov}(b)--(d) and the overlap distances and fractions in Figs.~\ref{fig:D_mirror_ext_lov}(e)--(g). This is useful for understanding the various changes in the free energy with respect to $\lambda$ in Fig.~\ref{fig:D_mirror_ext_lov}(a). Typically, there is a change in either the slope or the curvature of $F(\lambda)$ at each of the transition boundaries.

 In regime~(i) at $\lambda\lesssim -50$, the polymers do not overlap and are not in contact. Thus, the free energy as well as the extension lengths for the whole polymers and for the individual loops are invariant with respect to variations in $\lambda$. In regime~(ii) at $-50 \lesssim \lambda \lesssim 20$, the whole-polymer extension and internal-loop extension decreases as $\lambda$ increases and the polymer centers come closer together. As noted above, this is the same situation as in regime~III for non-cross-linked polymers where the centers are close enough for the polymers to interact, but not sufficiently close to overlap, leading to longitudinal compression of the chains with a degree that increases with $\lambda$. Note that it is the pairs of small loops that are in contact with each other, as illustrated in Fig.~\ref{fig:D_mirror_ext_lov}(i). At $\lambda\approx -22$, it becomes more entropically favorable for the polymers to assume a state of mutual overlap. This is analogous to the II-III transition in Fig.~\ref{fig:archA_regimes}. Initially, there is mainly overlap between the pairs of small loops, leading to a spike in the small-loop extension in Fig.~\ref{fig:D_mirror_ext_lov}(d). As evident in Fig.~\ref{fig:D_mirror_ext_lov}(g), the fraction of each small loop overlapping with a large loop rapidly increases until they are completely nested in the large loop at $\lambda\approx -10$.  Regime~(iii) could be further subdivided into regimes where small rings partially overlap with each other but either do or do not partially overlap the large loop. However, the point where the small loops just begin to overlap with the large loop appears to happen very close to the value of $\lambda$ where pairs of small loops begin to partially overlap with the other small rings. This is likely a result of the fact that it is much less entropically costly for the small rings to overlap with the larger ring than with other small rings.  This combination of transitions appears to happen over a very narrow range of $\lambda$; consequently, we do not find it useful to distinguish between these two sub-regimes. 
 
 At $\lambda\approx -10$, regime~(iii) transitions to regime~(iv), at which point the large loops on the polymers begin to overlap. In addition, regime~(iv) is characterized by small loops that are completely nested within the larger loops of the other polymer, as evident in Fig.~\ref{fig:D_mirror_ext_lov}(g). The free energy  as well as the extension lengths of the whole polymers and of the large loops increase slightly here as the degree of polymer overlap continues to grow with $\lambda$. Note that the point where the polymer centers of mass coincide ($\lambda=0$) happens to lie in this domain. At $\lambda\approx 4$ the system move into regime~(v), a transition which is characterized by an abrupt reduction in the whole-polymer and large-loop extension lengths, as a consequence of the increased lateral freedom provided to the short rings. Regime~(v) is characterized by small loops that only partially overlap with the larger loops. This is also where the whole-polymer and large-loop extensions plateau, as does the whole-polymer overlap. At $\lambda\approx 18$, there is a second abrupt decrease in the various extension lengths, and the system enters into regime~(vi), characterized by plateaus in the whole-polymer and large-loop extension lengths. Here, the small loops are fully segregated from the large loops, which are themselves partially overlapping. At $\lambda\approx 30$, the whole-polymer and large-loop extension lengths abruptly decrease to values well below the values for isolated polymers. Here, the system enters regime~(vii), characterized by fully segregated polymers that are in contact and compressed. This is analogous to regime~(ii), except that in regime~(vii) it is the large loops that are pressed up against each other, whereas in regime~(ii) it is the small loops. As $\lambda$ increases, the polymers become less compressed, leading to an increase in the extension lengths. Concurrently, the overlap between the polymers is near constant at zero. Finally, at $\lambda\approx 60$, the system transitions to regime~(viii), characterized by segregated and separated polymers. As in regime~(i), the extension lengths and the free energy are unaffected by $\lambda$. 

Let us now consider the case of two Arch-D chains arranged in the parallel orientation along the channel. Figure~\ref{fig:D_nonmirror_ext_lov} shows a collection of results characterizing the segregation behavior of the system. Unlike the case for the antiparallel polymer orientation, the extension lengths and the overlap distances differ between the two polymers at any given $\lambda$. Note that the extension length and overlap for the two polymers are related such that the value for polymer 1(2) at $\lambda$ is equal to that for polymer 2(1) at $-\lambda$. In addition, $F(\lambda) = F(-\lambda)$. Because of these symmetries, we need only plot results for $\lambda\geq 0$. 

We identify four different segregation regimes. At $\lambda=0$, the two polymers are equivalent, and the extension and overlap distances are equal. As $\lambda$ increases, the symmetry is broken, and the extension lengths between the two polymers  become markedly different, as evident in Figs.~\ref{fig:D_nonmirror_ext_lov}(b)--(d). We label the two polymers 1 and 2 and define $\lambda=Z_{{\rm cm},1} - Z_{{\rm cm},2}$, where $Z_{{\rm cm},i}$ is the center of mass of polymer $i$. Interestingly, at low values $\lambda$ polymer 2 is often completely nested within the large loop of polymer 1 most of the time, as is clear from Fig.~\ref{fig:D_nonmirror_ext_lov}(h). For another measure that corroborates this interpretation, we note that at $\lambda\approx 7$, where polymer 1 is most extended, the mean extension length of the large loop of polymer 1 ($\approx 48$) just slightly exceeds the full extension length of polymer 2  ($\approx 47$). At this $\lambda$, a small portion of the small chains of polymer 1 still overlap with  polymer 2 (Fig.~\ref{fig:D_nonmirror_ext_lov}(g)). We label this regime~(i) and illustrate its basic features in  Fig.~\ref{fig:D_nonmirror_ext_lov}(i). Such whole-polymer nesting was absent for any regime for the antiparallel polymer orientation. In this regime, the difference of each extension length of the two polymers is greater than for any separation distance outside this regime, with the nested polymer (2) being considerably shorter than the outer polymer (1). While the free energy function for polymers without cross linking tends to be constant in the nesting regime, the asymmetry of these molecules causes $F$ to decrease with $\lambda$. This is a result of the a reduction in the inter-polymer overlap of the short loops which the system avoids because it is more entropically costly (i.e. lower entropy) than overlap between small and large loops, which leads to reduced crowding. 

Additional insight into the system behavior in regime~(i) is provided by Fig.~\ref{fig:D_nonmirror_ext_lov}(j). In these calculations, the centre-of-mass distance $\lambda$ is constrained to lie within a very narrow range centered at values $\lambda_0$, which were chosen to be close to $\lambda=0$. The free energy was measured with respect to $\zeta\equiv Z_{{\rm ext},1}-Z_{{\rm ext},2}$, the difference in the two extension lengths. At $\lambda_0=0$, there is a broad flat region centered at $\zeta=0$, indicating an indifference to the degree of nesting (including the case of no nesting for $\zeta=0$). This contrasts with the behavior of systems of linear polymers and ring polymers without cross links\cite{polson2018segregation, polson2021free} where a clear preference for nesting is evident from a free-energy barrier at $\zeta=0$. Nesting becomes more likely at larger $\zeta$, which, as evident in Fig.~\ref{fig:D_nonmirror_ext_lov}(h) is about 25\% at $\lambda=0$. As $\lambda_0$ increases slightly, the free energy develops a clear minimum located at $\zeta>0$, where polymer 1 is longer than polymer 2. This free-energy feature promotes nesting, whose probability initially increases with $\lambda$, consistent with the results of Fig.~\ref{fig:D_nonmirror_ext_lov}(h) in the low-$\lambda$ portion of regime~(i). The importance and impact of topological modifications is highlighted by the contrasting nesting behavior at low $\lambda$ of the Arch-D polymers of this study and the simple ring polymers and linear polymers of Refs.~\onlinecite{polson2018segregation, polson2021free}. 

At $\lambda\approx 12$, the system transitions to regime~(ii), characterized by only partial overlap between the polymers (thus a cessation of whole-polymer nesting). Here, a portion of the large loop of polymer 2 and portions of the small loops of polymer 1 lie outside the overlap regime. Notably, the small loops of polymer 2 are themselves still completely nested within polymer 1. As $\lambda$ increases, the polymer overlap lessens, and thus the free energy decreases. The whole-polymer and main-loop extension length of both polymers tend to decrease slightly as more of the large loops enter the non-overlap zone as $\lambda$ increases. Conversely, the small loops of polymer 2 become more extended (Figs.~\ref{fig:D_nonmirror_ext_lov}(c) and (f)) while nevertheless staying nested in the large loop of the other polymer (Fig.~\ref{fig:D_nonmirror_ext_lov}(g)). At $\lambda\approx 25$ there is an abrupt reduction in all extension lengths (Figs.~\ref{fig:D_nonmirror_ext_lov}(b)--(d)) accompanied by an equally abrupt reduction to near zero of the average overlap fraction between each loop of one polymer and the entire other polymer (Figs.~\ref{fig:D_nonmirror_ext_lov}(g). In addition, the overlap distance between the polymers drops to near zero, after which it decreases at a very slow rate with $\lambda$. Taken together, these results demonstrate that the polymers have entered into a regime, labelled regime (iii),  where they are segregated, but compressed and in contact with one another. Similar regimes were noted above in Fig.~\ref{fig:D_mirror_ext_lov} for polymers in the antiparallel orientation, and for Arch-A ring polymers in Fig.~\ref{fig:archA_regimes}. As in those cases, increasing $\lambda$ gradually decompresses the polymers and their internal loops. In addition, the curvature of $F(\lambda)$ has shifted from negative in regime~(ii) to positive in regime~(iii), for reasons that were explained in depth in Ref.~\onlinecite{polson2021free} for the much simpler case of linear polymers. At $\lambda\approx 55$, the polymers enter regime (iv) where they are now separated along the channel. As before, this regime is characterized by constant free energy and extension lengths and an overlap ``distance'' that is negative and decreases linearly with $\lambda$. 

To summarize, the results of Figs.~\ref{fig:D_mirror_ext_lov} and \ref{fig:D_nonmirror_ext_lov} reveal rich segregation behavior of topologically modified ring polymers. Rather subtle variations in the shape of the free-energy functions are closely connected with the inter-polymer entropic repulsion between the internal loops, an understanding of which is aided by the behavior of simple ring polymers shown in Fig.~\ref{fig:archA_regimes}. Comparison of the results of the two figures also reveal the effects of the relative orientation of topologically asymmetric molecules in situations where confinement tends to block switching between the two states. In addition to the striking difference in the behavior of the various overlap and extension lengths, there is a notable difference in the overlap free energies of $\Delta\Delta F\approx 5k_{\rm B}T$, which is about a 10\% difference. 

\begin{figure}[!ht]
\begin{center}
\includegraphics[width=0.45\textwidth]{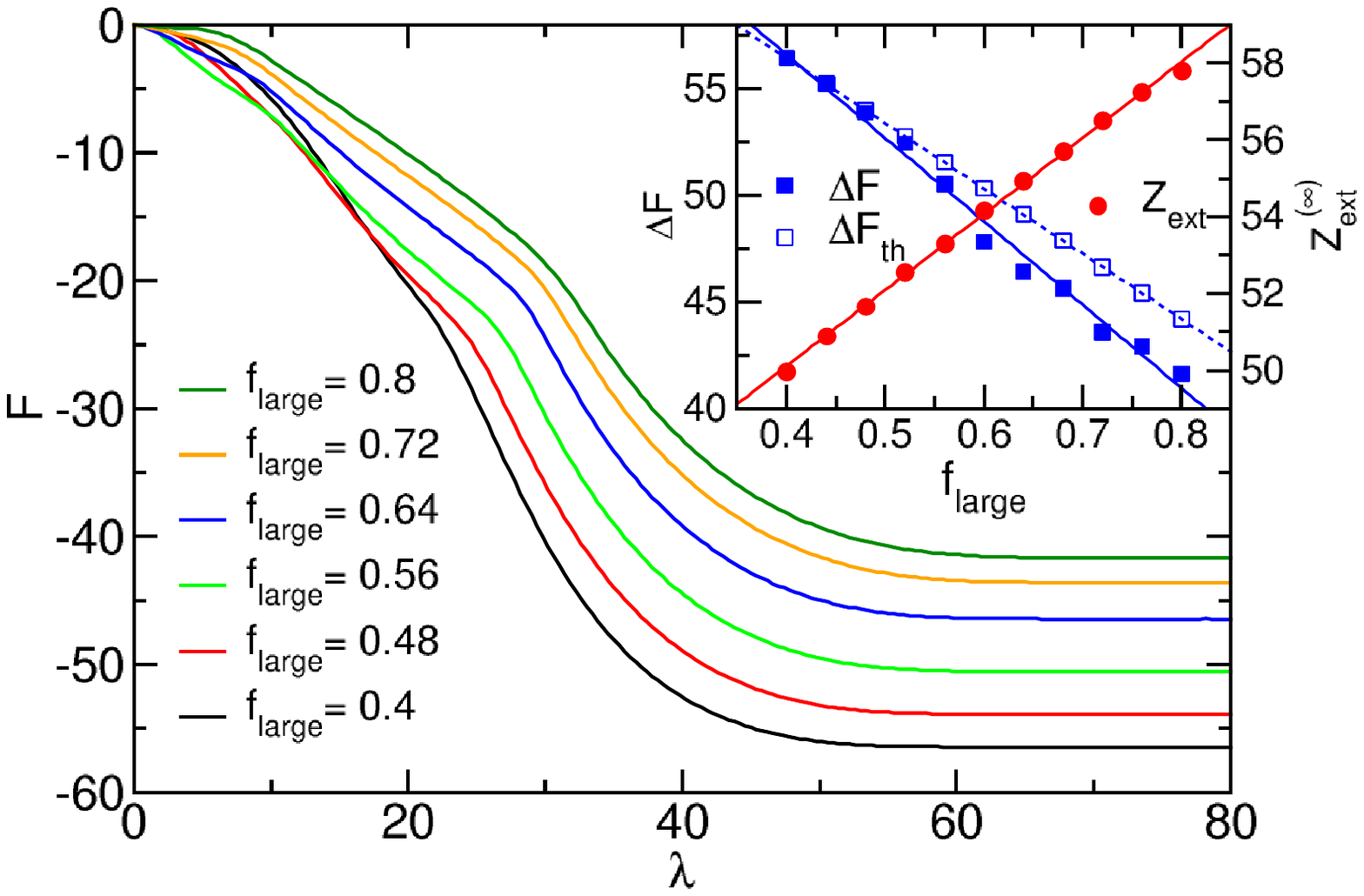}
\vspace*{-0.2in}
\end{center}
\caption{Free energy vs $\lambda$ for two parallel Arch-D chains in a channel of diameter $D=13$. Results are shown for different values of $f_{\rm large}$, the fraction of the total polymer contour length in the large loop. The inset shows the variation of the free-energy barrier height, $\Delta F$, and the large-$\lambda$-limit extension length, $Z_{\rm ext}^{(\infty)}$, with respect to $f_{\rm large}$. It also shows a theoretical prediction for the barrier height, $\Delta F_{\rm th}$, using the theoretical model described in the text.}
\label{fig:F_loop_ratio}
\end{figure}

It is of value to examine the effects of changing the structure of polymers of a given architecture in a manner  that also preserves their topology. We consider the case of Arch-D polymers where the binding site locations are varied in a manner so as to change the  relative contour length of the internal loops while preserving the symmetry of the structure. Figure~\ref{fig:F_loop_ratio} shows free energy functions for various values of $f_{\rm large}$, the fraction of the total contour length of the large loop. Two main trends are clear. First, the height of the free energy barrier, defined $\Delta F \equiv F(\lambda_{\rm max}) - F(\lambda=\infty)$, decreases monotonically as more of the polymer contour becomes contained in the large loop. The inset shows that the decrease is roughly linear with respect to $f_{\rm large}$. Second, the free energy levels off at higher values of $\lambda$ as $f_{\rm large}$ increases; consequently, the width of the free-energy barrier increases. These origin of these trends is straightforward. The small loops of the polymer tend to overlap along the channel in a region of higher local density than that occupied by the larger loop. Decreasing $f_{\rm large}$ increases the size of the small loops, thus drawing in a larger number of monomers to this higher density region. This has two effects on the overlap properties of the polymers. First, the higher the density, the more constricted the polymers are along the channel, i.e., the extension length is reduced. The inset of the figure shows that the large-$\lambda$ extension length, $Z_{\rm ext}^{(\infty)}$, decreases monotonically (and nearly linearly) with $f_{\rm large}$. This means that the polymers have to come closer together at lower values of $\lambda$ before they begin to overlap and the free energy rises, roughly where $\lambda\approx Z_{\rm ext}^{(\infty)}$. This explains the narrower free energy barrier for lower values of $f_{\rm large}$. In addition, the larger the size of the high-density regions of the small loops overlapping any portion of the other molecule, the greater the entropic penalty for overlap will be. This explains why the barrier height increases with decreasing $f_{\rm large}$.

A theoretical estimate of the free energy barrier height, $\Delta F\equiv F(\lambda=0)-F(\lambda=\infty)$, can be obtained using a theoretical model first proposed in Ref.~\onlinecite{jung2012ring}. Here, two polymer strands that overlap along the channel can be viewed as equivalent to two polymers confined to channels of half the cross-sectional area of the true channel. Thus, the diameter of the imaginary tubes is $D'=D/\sqrt{2}$, where $D$ is the diameter of the physical tube. A collection of $n$ overlapping strands can likewise each be viewed as effectively occupying a tube of diameter $D'=D/\sqrt{n}$. The de~Gennes blob model is employed to estimate the confinement free energy of each virtual linear polymer of length $N'$ using $\Delta F_{\rm c}=F_{\rm bl} N'(D')^{-1/\nu}$. Here, $\nu$ is the Flory scaling exponent for self-avoiding chains and  the blob free-energy prefactor, $F_{\rm bl}$, is of order $k_{\rm B}T$. The free energy is calculated for the case of two segregated polymers and for the case of two overlapping polymers, and $\Delta F$ is given by the difference of the two. A calculation for two Arch-D chains is presented in the appendix for arbitrary $f_{\rm large}$. The result is $\Delta F = F_{\rm blob} c(f_{\rm large}) ND^{-1/\nu}$, where the function $c(f_{\rm large})$ is given by Eq.~(\ref{eq:cfl}). The results of the calculation using $\nu=0.588$ are shown in the inset of Fig.~\ref{fig:F_loop_ratio}. As the value of $F_{\rm bl}$ is not known, we have arbitrarily chosen a value such that the value coincides with the simulation value at $f_{\rm large}=0.4$. The theory correctly predicts that $\Delta F$ decreases with increasing $f_{\rm large}$ and even reproduces the linear variation. However, the rate of decrease is noticeably lower for the theory, with $d(\Delta F_{\rm th})/df_{\rm large}=-30.1$ as compared to $d(\Delta F)/df_{\rm large}=-39.1$. Such a quantitative discrepancy is typical for this theoretical model and arises from the nature of the approximations employed therein, as we have noted in previous studies.\cite{polson2014polymer, polson2018segregation, polson2021free}

Let us expand our investigation of segregation of confined cross-linked polymer and examine the effects of varying the polymer architecture and varying the confinement dimension. Figure~\ref{fig:delFvsD} shows the variation of the free-energy barrier height, $\Delta F$, with respect to the confining cylinder diameter, $D$, over the range of $D=11-17$. Results for five different polymer architectures are shown. In most cases, the polymers are in the parallel configuration, though results are shown for Arch-D chains in both orientations. Two main results are apparent. First, for each architecture, the barrier height exhibits power-law scaling, with $\Delta F\propto D^{-\beta}$. The scaling exponent of the fits, shown as the dashed lines in the figure, are: $\beta=1.80\pm 0.01$ for Arch-A, $\beta=1.84\pm 0.01$ for Arch-B, $\beta=1.86\pm 0.01$ for Arch-C, $\beta=1.84\pm 0.02$ for Arch-D, and $\beta=2.03\pm 0.01$ for Arch-E. For the case of antiparallel Arch-D polymers, the exponent is $\beta=1.87\pm 0.01$. These are close to the the values measured in our previous studies on linear\cite{polson2014polymer, polson2021free} and ring\cite{polson2018segregation} polymers. The second notable feature in Fig.~\ref{fig:delFvsD} is the variation of the barrier height with polymer architecture, with a trend where $\Delta F$ generally increases with the complexity of the polymer architecture. In particular, with the exception of Arch-B, increasing the number of internal loops generally increases $\Delta F$. Though Arch-C and Arch-D each have three loops, the difference in the size and placement of the loops leads to appreciable differences in the barrier height.

Estimates of the free energy barrier height, $\Delta F\equiv F(\lambda=0)-F(\lambda=\infty)$, can be obtained using the theoretical model used in the analysis of the data in Fig.~\ref{fig:F_loop_ratio} above. Calculations for the other architectures are straightforward and were carried out following the same approach as that used for Arch-D polymers in the appendix. In each case, multiple overlapping strands are modelled as linear polymers confined to effective tubes of reduced diameter, and the de~Gennes blob-model is used to estimate the confinement free energy of each virtual linear polymer. As for Arch-D, simplifying assumptions are made regarding the internal and inter-polymer overlap of individual loops in the overlapping ($\lambda=0$) and segregated states (e.g. neglecting nesting effects).  For this reason, as well as for others noted in our previous studies,\cite{polson2014polymer, polson2018segregation, polson2021free} the theoretical model is only expected to provide a rough approximation for $\Delta F$. In all cases, it is predicted that $\Delta F_\alpha = c_\alpha F_{\rm bl} ND^{-1/\nu}$, where the numerical prefactor $c_\alpha$ depends on the specific architecture, where $\alpha$ denotes A, B, C, D and E. This result was previously reported in Ref.~\onlinecite{mitra2022topology}. The prediction that $\Delta F\propto D^{-\beta}$, where $\beta=1/\nu\approx 1.70$ is close to observed scaling with $\beta$ in the range of $1.8-2.0$ for various architectures. As in our earlier work,\cite{polson2014polymer, polson2018segregation, polson2021free} the measured scaling exponents are somewhat larger than the theoretical value. 

Rather than directly comparing the measured and theoretical values of the barrier height, it is more useful to employ the ratio $r_{\alpha A}\equiv \Delta F_\alpha / \Delta F_{\rm A}=c_\alpha/c_{\rm A}$. Such a quantity is independent of the unknown value of the blob free-energy prefactor $F_{\rm bl}$. Table~\ref{tab:table1} shows the measured and theoretical values for $r_{\alpha A}$ for each architecture. While the theoretical values are independent of $D$, there is a slight dependence on $D$ in the simulation data. In the case of Arch-B, the ratios are all within 2\% of the predicted value of unity. In this case, the two cross links do not lead to the formation loops that cause additional overlapping of polymer strands along the channel. The theoretical model correctly predicts that $1\approx r_{\rm BA} < r_{\rm CA} < r_{\rm DA} < r_{\rm EA}$. However, the individual predictions are not as accurate as for $r_{BA}$. For C and D, the theory overestimates the values by 8\% and 9\%, respectively, while for architecture~E, it underestimates it by about 10\%. As noted earlier, the magnitude of the quantitative discrepancies are consistent with results of previous studies and highlight the limitations of the theoretical model. Nevertheless, the qualitative trends of the simulation results are generally consistent with the predictions.

\begin{table}[h]
\begin{ruledtabular}
\begin{tabular}{l c c c c}
                    & $r_{\rm BA}$  & $r_{\rm CA}$  & $r_{\rm DA}$  & $r_{\rm EA}$  \\
                    \hline
Theory          & 1.00                & 1.16               & 1.39                & 1.44 \\
\hline        
Simulation & & & & \\         
D = 11 &  0.985 &   1.08 &  1.29 & 1.67 \\
D = 12 &  0.991 &   1.09 &  1.28 & 1.65 \\
D = 13 &  0.978 &   1.07 &  1.29 & 1.62 \\
D = 14 &  0.975 &   1.08 &  1.27 & 1.59 \\
D = 15 &  0.980 &   1.07 &  1.27 & 1.57 \\
D = 16 &  0.973 &   1.06 &  1.26 & 1.53 \\
D = 17 &  0.979 &   1.06 &  1.27 & 1.52 \\
\hline
Av. Relative Diff. & 0.020 & 0.081 & 0.090 & -0.096
\end{tabular}
\end{ruledtabular}
\caption{Free-energy barrier height ratio, $r_{\alpha {\rm A}}\equiv \Delta F_\alpha / \Delta F_A$, where $\alpha$ denotes the architecture label B, C, D and E. Simulation results are shown for various values of the cylinder diameter. The final row is the average relative difference between the theoretical value and the values of the simulation data. Positive (negative) values signify theoretical over(under)estimates. Theoretical predictions are calculated using the blob model, as described in Ref.~\onlinecite{mitra2022topology}.}
\label{tab:table1}
\end{table}


\subsection{Finite-length channels}
\label{subsec:finite}

\begin{figure*}[!ht]
\begin{center}
\vspace*{0.0in}
\includegraphics[width=0.85\textwidth]{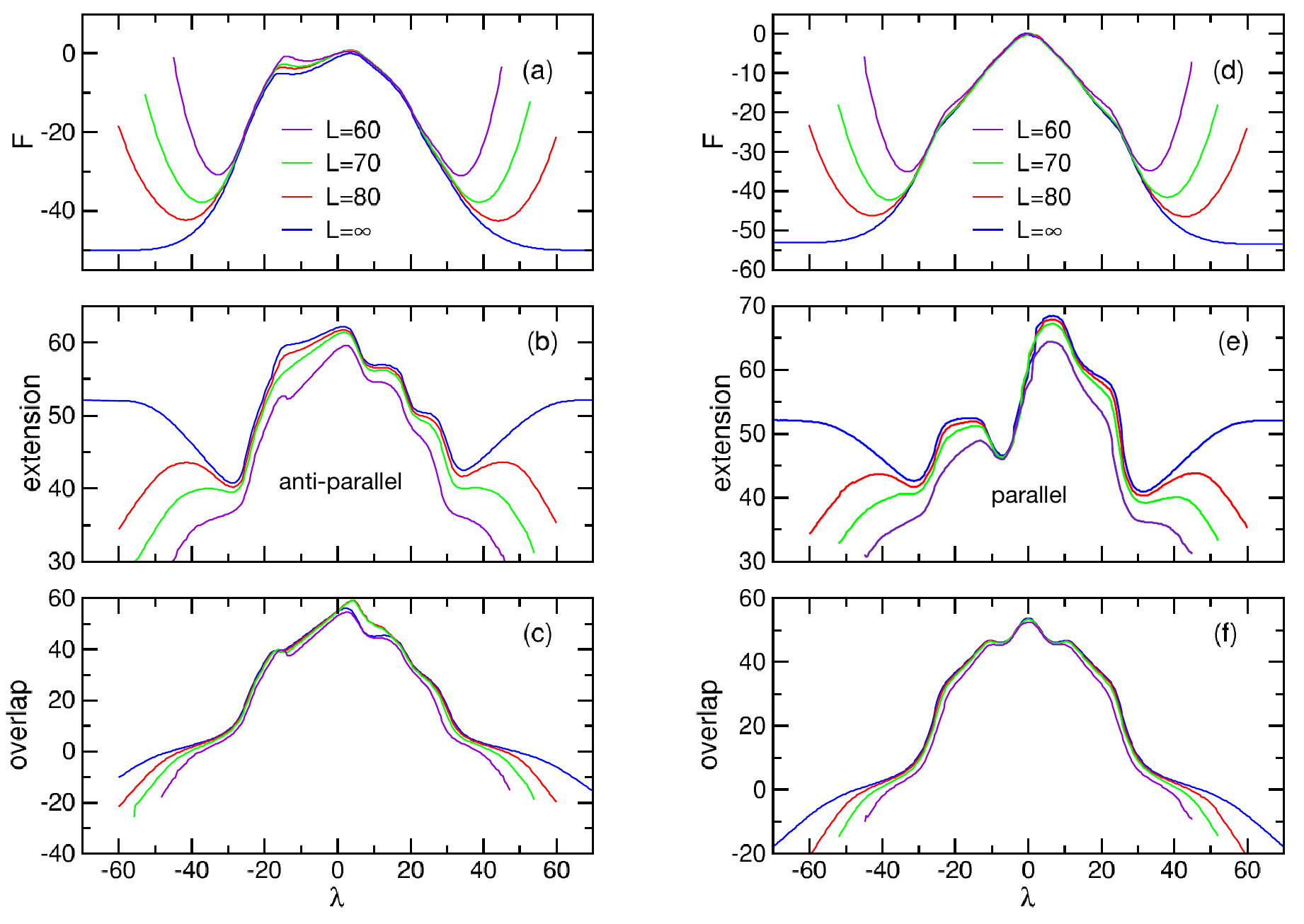}
\vspace*{-0.2in}
\end{center}
\caption{Effects of variation of the cavity length for chains of architecture D confined to a finite-length cavity of diameter $D=13$ for both antiparallel and parallel configurations. (a) Free energy, (b) extension length of the polymers, and (c) overlap length of the polymers in the antiparallel configuration. Panels (d)--(f) show analogous results for the parallel configuration. For the extension curves shown in (e), results are shown only for polymer 1, defined in Fig.~\ref{fig:D_nonmirror_ext_lov}(i). Results for polymer 2 are given by those for polymer 1 upon performing the  transformation $\lambda\rightarrow -\lambda$.}
\label{fig:F_ext_lov_L}
\end{figure*}

Consider now the effects of longitudinal confinement in which the chain segregation occurs in cylinders of finite length. Figure~\ref{fig:F_ext_lov_L} shows the results for the free energy and whole-chain extension and overlap of Arch-D chains in such finite-length cavities. Figures~\ref{fig:F_ext_lov_L}(a)--(c) show results for chains in the anti-parallel orientation, while Figs.~\ref{fig:F_ext_lov_L}(d)--(f) show analogous results for parallel chains. In each case, results are shown for cylinder lengths of $L=60, 70$ and $80$, as well for infinite-length cylinders. The values of $L$ are chosen so that the chains feel the presence of the cylinder end caps before segregation is complete. This results in free-energy functions that peel away from the $L=\infty$ curves as $\lambda$ increases in magnitude on either side of the central barrier (Figs.~\ref{fig:F_ext_lov_L}(a) and (d)). The rapid increase in $F$ at large $|\lambda|$ is a result of the loss of entropy associated with the polymer pressing against the caps. The resulting free energy minima are located at positions corresponding to the most probable centre-of-mass positions of the polymers. As expected, this distance decreases as the confinement length decreases. Also notable is the reduction the barrier height, $\Delta F\equiv F(\lambda_{\rm max})-F(\lambda_{\rm min})$, with increasing confinement. At the same time, the free energy for states with a high degree of polymer overlap is barely affected by longitudinal confinement. In this case, the combined extension length of the two polymers, $L_{\rm tot}\equiv 2Z_{\rm ext}-Z_{\rm ov}$, satisfies $L_{\rm tot}\lesssim L$, and the free-energy curves mostly overlap. It is interesting to note that, while the $F(\lambda)$ functions are not symmetric for the anti-parallel case, the free-energy barrier heights measured with respect to each minimum are essentially equal for each value of $L$.

Like the free energies, the extension and overlap lengths for finite $L$ also track closely to those for $L=\infty$ for both parallel and antiparallel orientations, with significant deviations only in larger-$|\lambda|$ regions where the polymers are pressed up against the end caps. There is a somewhat greater deviation of the extension length in the low-$|\lambda|$ regions for the shortest cavity, $L=60$.

\begin{figure}[!ht]
\begin{center}
\vspace*{0.0in}
\includegraphics[width=0.45\textwidth]{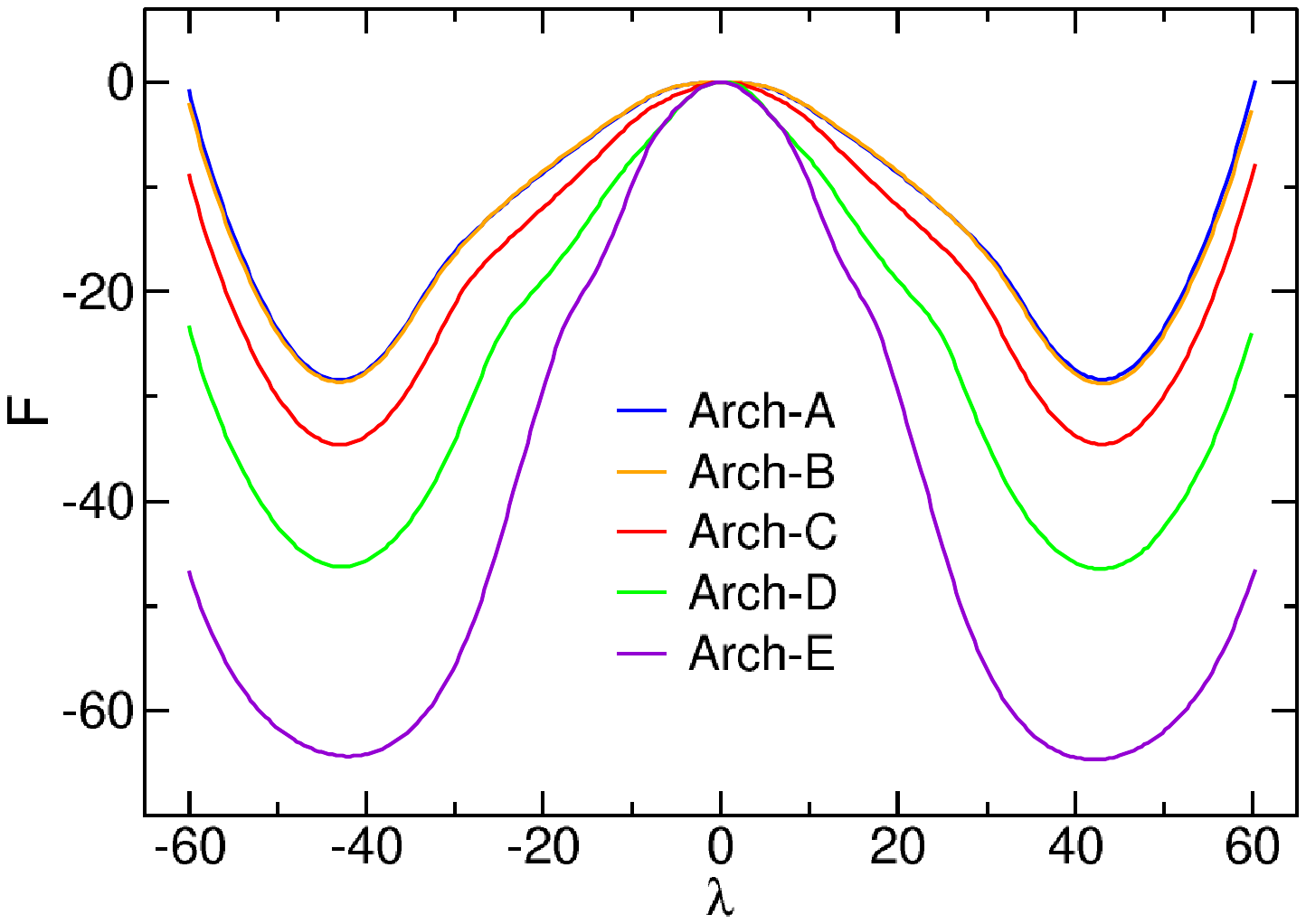}
\vspace*{-0.2in}
\end{center}
\caption{Free-energy functions for different polymer architectures for polymers in the parallel configuration confined to a finite-length channel of length $L=80$ and diameter $D=13$.  }
\label{fig:all_arch_F}
\end{figure}

Figure~\ref{fig:all_arch_F} shows free-energy functions for all five architectures for the case of parallel-orientation chains confined to a cylinder of diameter $D=13$ and length $L=80$. The free-energy barrier heights are related by  $\Delta F_{\rm A} \approx \Delta F_{\rm B} <  \Delta F_{\rm C} <  \Delta F_{\rm D} <  \Delta F_{\rm E}$, consistent with the results for infinite-length cylinders in Fig.~\ref{fig:delFvsD}. The position of the minimum is not much affected by varying the architecture, with only a small shift to lower $\lambda$ observed for Arch-E. This is a result of the shorter extension length of this polymer. There is a noticeable shift to smaller $\lambda$ of the shoulders of the barrier, features that are expected to depend on the number and placement of internal loops, as discussed in detail in the context of Figs.~\ref{fig:D_mirror_ext_lov} and \ref{fig:D_nonmirror_ext_lov} for Arch-D chains. Also noticeable is the widening of the free energy minima, particularly so for Arch-E. Again, this is a result of the shorter extension length for these chains. When the infinite-$L$ extension length $Z^{(\infty)}_{\rm ext}$ satisfies $2Z^{(\infty)}_{\rm ext} \lesssim L$, then there is space for separation of the polymers after they segregate before they push up agains the end caps, resulting in a wider minimum. The same effect is evident in Figs.~\ref{fig:F_ext_lov_L}(a) and (d) for Arch-D chains upon increasing $L$. 

It is useful to compare further the effects of varying architecture on the free-energy barrier height for the infinite- and finite-length cases. This will provide a means to evaluate the validity of using the de~Gennes blob model to estimate the overlap free energy for finite-length cavities using a theoretical model designed for open channels. Such estimates were employed in Ref.~\onlinecite{mitra2022topology} as well as in the supplemental information of Ref.~\onlinecite{harju2024loop}. It is expected that the predictions will be reasonably good provided the longitudinal confinement is not too strong. Table~\ref{tab:table2} compares the barrier height, $\Delta F$, for the cases of $L=80$ and $L=\infty$. In addition, we also compare the ratio $r_{\alpha {\rm A}} \equiv \Delta F_\alpha / \Delta F_{\rm A}$, which is an estimate of the ratio of the number of blobs for a given architecture relative to that for Arch-A. As noted earlier, this quantity has the convenient property of being independent of the unknown value of the free energy per blob. In accord with the theoretical prediction, $r_{\alpha {\rm A}}$ increases for architectures C, D and E in that order for both $L=\infty$ and $L=80$. This trend is due to the increase in the number of blobs relative to open rings (Arch-A) arising from the forced overlap of strands associated with the internal loops. Note, however, that ratio increases much faster for $L=80$ than for $L=\infty$.  For Arch-D and Arch-E chains, the discrepancy with theory is 15\% and 37\%, respectively, for $L=80$, compared to 8\% and 12\% for $L=\infty$. The worsening accuracy of the predictions for the more topologically complex polymers at finite $L$ is likely a consequence of the increase in volume fraction associated with longitudinal confinement. Such an increase likely pushes the system into the semi-dilute regime where there are expected to be a greater number of smaller-sized blobs.\cite{jun2010entropy} In this situation, the picture of overlapping strands occupying effective tubes likely begins to break down, leading to barrier heights $\Delta F$ that are even further removed from the theoretically predicted values than those listed for $L=\infty$ in Table~\ref{tab:table2}. We conclude that the blob-model approach becomes significantly less accurate as longitudinal confinement increases.

\begin{table}[h]
\begin{ruledtabular}
\begin{tabular}{l c c c c c}
       &     A & B & C & D & E \\ \hline
$\Delta F ~(L=\infty)$    &  41.8 & 40.8 & 44.8 &  54.0 & 67.4 \\
$\Delta F ~(L=80)$        &  28.4 & 28.8 & 34.6 & 46.5 & 64.7  \\ \hline
$r_{\alpha {\rm A}}  ~(L=\infty)$ &  1     &  0.98 & 1.07  &  1.29 &  1.64 \\
$r_{\alpha {\rm A}}  ~(L=80)$     &  1    &  1.01 & 1.22  &  1.64 &   2.28 \\ 
$r_{\alpha {\rm A}}$ (theory)      & 1     &  1     & 1.16  &   1.39 &   1.44\\
\hline
Rel. Diff. ($L=\infty$)                  & 0     &  0.02 &  0.084 &   0.078 &   -0.122 \\
Rel. Diff. ($L=80$)                      & 0     &  -0.01 & -0.049 &  -0.152 &   -0.368 
\end{tabular}
\end{ruledtabular}
\caption{Free-energy barrier height data for polymers of different architectures. Results are shown for confinement in a cylinder of diameter $D=13$ for infinite-length cylinders and for capped cylinders of length $L=80$. The table shows the barrier heights, $\Delta F \equiv F(\lambda_{\rm max})-F(\lambda_{\rm min})$, the barrier height ratios, defined $r_{\alpha {\rm A}} \equiv \Delta F_\alpha/\Delta F_{\rm A}$ for  architecture $\alpha$, as well as the relative difference in the ratios, defined $(r_{\alpha {\rm A}}({\rm theory})-r_{\alpha {\rm A}})/r_{\alpha {\rm A}}$. Positive (negative) values signify theoretical over(under)estimates. Theoretical predictions are calculated using the blob model, as described in Ref.~\onlinecite{mitra2022topology}. }
\label{tab:table2}
\end{table}

\section{Conclusions}
\label{sec:conclusions}

In this study, we have employed MC simulations to characterize the effects of internal cross linking  on the segregation of two identical ring polymers confined to cylindrical spaces of infinite and finite length. As in our previous studies,\cite{polson2014polymer, polson2018segregation, polson2021free} we focus on characterizing the behavior of the segregation free energy functions, where the configurational free energy is measured with respect to the inter-polymer center-of-mass distance along the cylinder axis. The polymer architectures were selected to highlight the effects of varying the number, size and placement of the internal loops arising from the cross linking. Two architectures (Arch-D and Arch-E) were selected because of their prominence in several recent molecular dynamics studies of polymer segregation.\cite{mitra2022polymer, pande2024topology, bhandarkar2026entropic, doshi2026kinetics} As in our previous work on linear and ring polymers, we also examine the effects of varying the dimensions of the confining cylinder. To better understand the free-energy functions, we also measure and analyze the extension and overlap lengths for the internal loops as well as for the entire polymers. One key goal of this study is a quantitative assessment of the accuracy of a  theoretical model based on the de~Gennes blob model that has been used to estimate both the height and shape of the segregation free-energy functions and to help interpret the results of segregation dynamics simulations.  

As in the case of simple linear and ring polymers, the segregation of cross-linked polymers can be characterized by distinct stages. However, the presence of the internal loops increases the complexity of the process and, consequently, the number of stages. An additional complication is that when the cylindrical confinement is sufficiently high, two cross-linked polymers with asymmetric topologies tend adopt two distinct relative orientations, interconversion between which is separated by a free-energy barrier that grows with decreasing cylinder diameter. The observed stages of segregation differ between the two orientations. However, the free-energy barriers tend to be comparable in height and width, with roughly a 10\% difference in the barrier height observed for Arch-D chains. Thus, while flipping between states during segregation may be possible, doing so is unlikely to lead to any significant change in the resulting entropic force. 

The free-energy functions change significantly with the polymer topology. In accord with trends observed in previous simulations tracking the dynamics of segregation, the  barrier height (i.e. the free-energy cost of fully overlapping the polymers) tends to increase with increasing topological complexity. Essentially, this means that the entropic force increases with increasing number of internal loops and more asymmetric placement. For example, Arch-E chains, with five internal loops, have a notably higher barrier than the three architectures (B, C and D) with three internal loops. Further, chains with ``side-looped'' topologies (C and D) have a higher barrier heights compared to that for Arch-B, which has ``linearly looped'' structure.\cite{mitra2022topology} The quantification of the impact of molecular architecture on the free-energy functions is the most significant contribution of the present study.

Another focus of investigation was testing the accuracy of a modified version of the blob model that has been used in a number of studies to estimate the segregation free energy. In one previous study,\cite{polson2018segregation} we examined the validity of the model for both linear and ring polymers and found that persistent quantitative inaccuracies do arise from the underlying approximations but that the model still captures the main trends. In the present study, we report similar findings. For all polymer architectures, the free energy barrier height scales with cylinder diameter as $\Delta F \sim D^{-\beta}$ with $\beta\approx 1.8-2.0$, close to the model prediction of $\beta=1/\nu \approx 1.70$. The theoretical model accurately predicts the observed ordering of the proportionality coefficients for the various architectures, though the values are typically off by roughly 10\%. Imposing longitudinal confinement by capping the cylinders tends to reduce $\Delta F$, through the free-energy functions tend to remain unchanged from the $L=\infty$ case for high overlap. The blob-model prediction for $\Delta F$ tends to worsen in accuracy as the cylinder length is reduced, particularly so for more topologically complex chains.

Although this study was in part motivated by its relevance to chromosome segregation in bacteria, the model and methods have a number of features that limit its direct applicability to that phenomenon. First, a free-energy function is an equilibrium property, whereas chromosome segregation is an inherently out-of-equilibrium process. Nevertheless, in cases where entropy contributes appreciably to chromosome segregation, the segregation rate may still correlate with the entropic force associated with the equilibrium functions. In our previous work, we showed that the segregation kinetics proceeded quasi-statically for most of the process, deviating only at times where there was an abrupt change in the free energy.\cite{polson2021free} A second important deficiency of our model is the fact the fact that it considers the segregation of fully replicated polymers. In bacterial chromosomes, by contrast, replication and segregation are concurrent phenomena. Indeed, it is because the rate of replication occurs on timescales fast relative to the chromosome conformational relaxation that the system is driven out of equilibrium. A third limitation is the use of cross links at {\it fixed} locations along the chain, a feature justified by {\it in vivo} experimental results involving colocalization of {\it rrn} operons in {\it E. coli}. However, recent experimental and theoretical studies have provided considerable evidence that {\it dynamical} cross links created by SMC complexes help drive chromosome segregation. Finally, the coarse-grained model polymers employed in the present study were topologically constrained so that all loops remained permanently uncatenated. By contrast, bacterial chromosomes employ topoisomerase enzymes to facilitate strand passage and help resolve entanglements during segregation. This suggests that any internal loops may catenate and de-catenate repeatedly. Relaxing this topological constraint to permit such events will likely impact the free energy functions. 

In future work, we will address some of these limitations by measuring and characterizing the segregation free energies of partially replicated model chromosomes designed to permit chain crossing. In addition, cross linking will be implemented using models chosen to better resemble the structural features expected from the actions of SMC complexes. An example of this is provided by a recent simulation study of SMC loading at the {\it ori} sites of replicated arms of a model chromosome.\cite{harju2024loop} Finally, dynamics simulations will be used to complement the free energy calculations to quantify the impact of out-of-equilibrium behavior and so determine the degree to which equilibrium free energy functions are relevant to the process.

\begin{acknowledgments}
This work was supported by the Natural Sciences and Engineering Research Council of Canada (NSERC). We are grateful to the Digital Research Alliance of Canada and the Atlantic Computational Excellence Network (ACEnet) for use of their computational resources. We would like to thank Sheldon Opps for helpful discussions.
\end{acknowledgments}

\section*{Data Availability Statement}
The data that support the findings of this study are openly available in Zenodo at https://doi.org/10.5281/zenodo.21499558

\appendix

\section{Flipping free energy for Arch-D chains}
\label{app:a}

\begin{figure}[!ht]
\begin{center}
\vspace*{0.0in}
\includegraphics[width=0.45\textwidth]{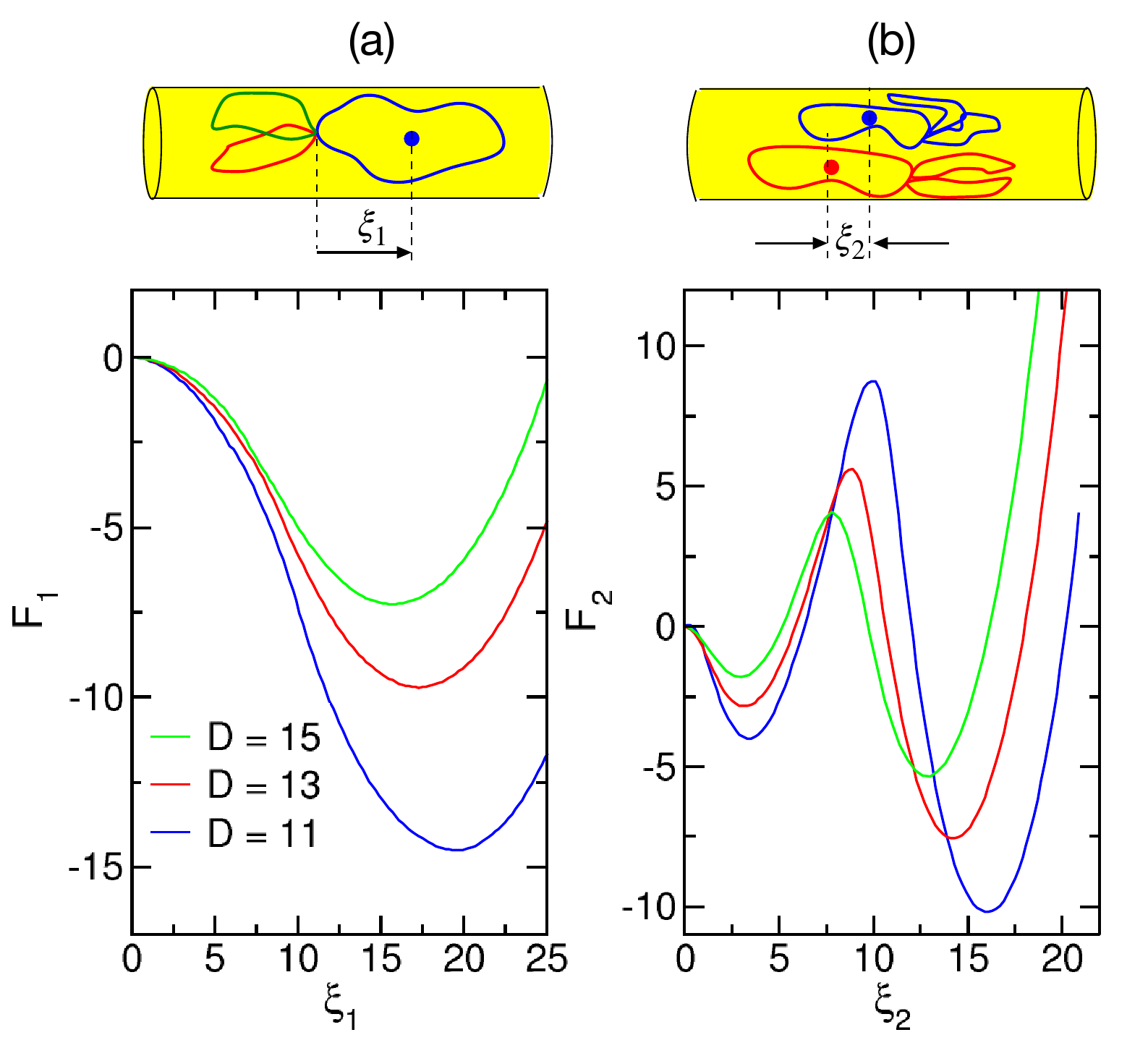}
\vspace*{-0.2in}
\end{center}
\caption{(a) Illustration of the definition of $\xi_{1}$ for a single Arch-D polymer. The dot denotes the center of mass for the large loop. The graph shows the variation of the free energy with respect to $\xi_1$. Results are shown for $D=11, 13$ and $15$. (b) Illustration of the definition of $\xi_{2}$ for two Arch-D polymers. The dots are the centers of mass of the two large loops. The graph shows the variation of the free energy with respect to $\xi_2$ for the same values of $D$ as in (a). While the plots only show $\xi_i\geq 0$, the case of $\xi_i< 0$ follows from the property that $F_i(-\xi_i)= F_i(\xi_i)$.  In both (a) and (b) polymers are confined to an infinitely-long cylinder. }
\label{fig:F_flip}
\end{figure}

Arch-D polymers confined to a channel tend to be configured with the large loop on side and the two small loops on the other along the channel.\cite{bhandarkar2026entropic} A single polymer orients with equal probability in either direction and can flip between the two orientations by overcoming a free-energy barrier. Likewise, two overlapping Arch-D polymers may be oriented in the same direction (the ``parallel" configuration) or in the opposite direction (the ``antiparallel" configuration), and flipping between the states requires overcoming a free energy barrier. In this appendix we calculate the barrier separating these pairs of states for isolated and for overlapping Arch-D polymers. To do this, we calculate the free energy as a function of a suitably chosen reaction coordinate. For the single polymer case, we choose this to be $\xi_1$, the displacement along the channel between center of mass of the large loop and the linking site where the small loops attach to the large loop. Similarly, for the overlapping polymer pair, we choose $\xi_2$, the displacement between the centers of mass of the large loops along the channel. The quantities are illustrated in Fig.~\ref{fig:F_flip}.

The functions $F_1(\xi_1)$ and $F_2(\xi_2)$ are calculated using the SCH method employed to calculate the overlap free energy function $F(\lambda)$, described in detail in Section~\ref{sec:methods}. In this case, window potentials of the form in Eq.~(\ref{eq:winPot}) are employed using the quantities $\xi_1$ and $\xi_2$ in place of $\lambda$. In the case of the overlapping pair, the centers of mass are constrained to lie very close together along the channel axis within the range of $[-\sigma,+\sigma]$.  This condition is essentially the same as that used to calculate $F(\zeta)$ in Fig.~\ref{fig:D_nonmirror_ext_lov}(j). Free-energy functions for $F_1(\xi_1)$ are shown in Fig.~\ref{fig:F_flip}(a), while those for $F_2(\xi_2)$ are shown in  Fig.~\ref{fig:F_flip}(b). Results are shown for three values of cylinder diameter, $D$. Since the curves are each symmetric about $\xi_i=0$, we show results only for $\xi_i\geq 0$.

For the single-chain system, the free energy has a minimum at a value $\xi_1^{\rm (min)}$ that increases with increasing confinement, a reflection of the greater longitudinal stretching of the polymer as $D$ decreases. A similar minimum (not shown) occurs at $-\xi_1^{\rm (min)}$, and a barrier with a maximum at $\xi_1=0$ separates them. As the channel narrows, the free energy barrier height $\Delta F_1\equiv F_1(0)-F_1(\xi_1^{\rm (min)})$ increases. The values are listed in Table~\ref{tab:table3}. At $\xi_1=0$, the internal loops all overlap, and the resulting monomer density increases with decreasing channel width. This substantially decreases the conformational entropy and leads to the observed in increase in barrier height with decreasing $D$.

\begin{table}[h]
\begin{ruledtabular}
\begin{tabular}{l c c  c c c c}
& \multicolumn{2}{ c }{$F_1$} & & \multicolumn{3}{ c }{$F_2$} \\
\cline{2-3} \cline{5-7}
                    & $\Delta F_1$  & $\xi_1^{\rm (min)}$  & & $\Delta F_{\rm min}$  & $\Delta F^{\rm (p)}$ & $\Delta F^{\rm (a)}$  \\
\hline         
D = 11 &  14.5   &   19.5 &&  6.20 & 12.75 & 18.95 \\
D = 13 &  9.7   &   17.3 && 4.70 & 8.46 & 13.16 \\
D = 15 &  7.3 &   15.7 && 3.54 & 5.87 & 9.41 \\
\end{tabular}
\end{ruledtabular}
\caption{Quantities characterizing the free-energy functions in Fig.~\ref{fig:F_flip}. $\Delta F_1$ is the free-energy barrier height for $F_1(\xi_1)$. $\xi_1^{\rm (min)}$ is the position of the minimum of $F_1(\xi_1)$. $\Delta F_{\rm min}$ is the difference in $F_2$ at the two minima. $\Delta F^{\rm (p)}$ is the barrier height for $F_2$ measured from the parallel-configuration minimum, and $\Delta F^{\rm (a)}$ is the barrier height measured from the antiparallel-configuration minimum. }
\label{tab:table3}
\end{table}

The functions $F_2(\xi_2)$ of Fig.~\ref{fig:F_flip}(b) are characterized by two free energy minima separated by a barrier. The lower-$\xi_2$ minimum corresponds to the parallel configuration, while the higher-$\xi_2$ barrier corresponds to the antiparallel case. Interestingly, the large loops for parallel polymers are slightly displaced from one another, with a slight free-energy barrier at $\xi_2=0$. This is reminiscent of the overlap free-energy barrier for both ring and linear polymers\cite{polson2018segregation, polson2021free} characterizing the nesting state for $\lambda\approx 0$. Here, it is likewise due to a nesting process, one involving only the two large loops. More noteworthy is the fact that the free-energy minimum for the anti-parallel state, $F_{\rm min}^{(p)}$, is lower  than that for the parallel state, $F_{\rm min}^{(a)}$ with a difference of $\Delta F_{\rm min} \equiv F_{\rm min}^{(a)} - F_{\rm min}^{(p)}$  that grows with increasing confinement. In addition, the free-energy barrier also grows as $D$ decreases. Values of the barrier height relative to the parallel-configuration minimum, $\Delta F^{\rm (p)}$, and that relative to the antiparallel-configuration mimimum, $\Delta F^{\rm (a)}$, are listed in Table.~\ref{tab:table3}. The increase in $\Delta F^{(i)}$ with decreasing $D$ is due to the increase in monomer crowding of due to overlapping internal loops in the transition state, which in turn significantly decreases the polymer conformational entropy.

\section{Theoretical estimate of $\Delta F$ for Arch-D polymers of arbitrary large-loop size}
\label{app:b}

In this appendix, we present a theoretical calculation used to estimate the free-energy barrier height of two cylinder-confined Arch-D polymers of length $N$ arranged in the parallel configuration. Rather than limit the calculation for the specific binding sites illustrated in Fig.~\ref{fig:archpol_illust}, we consider the general case where the large-loop size has contour length $f_{\rm large}N$ and the small loops each have length $(1-f_{\rm large})N/2$. The calculation is very close to that presented in the appendix of Ref.~\onlinecite{mitra2022topology}. 

\begin{figure}[!ht]
\begin{center}
\vspace*{0.0in}
\includegraphics[width=0.45\textwidth]{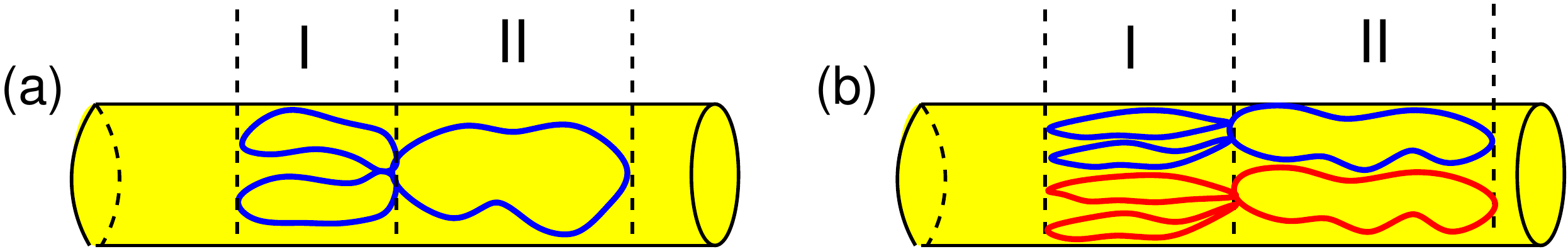}
\vspace*{-0.2in}
\end{center}
\caption{Illustration of the definitions of regions I and II in the text. (a) The two overlapping small loops in region~I can be modelled as four overlapping linear polymers of half the loop contour length. Likewise, the single large loop is similarly modelled as two overlapping linear polymers. (b) Two overlapping polymers. In this case, the four small loops in region~I are analogous to eight overlapping linear polymers, while the large loops are analogous to 4 overlapping linear strands. }
\label{fig:appendix_model}
\end{figure}

Consider the case of a single isolated Arch-D polymer in a channel of diameter $D$. Assume the two small polymers overlap with each other with equal extension length. Call this section of the channel region I. Here, each small loop is a ring polymer, and the two overlapping loops can be viewed as the equivalent of four overlapping linear polymers, each of length $N'=(1-f_{\rm large})N/4$. Following the suggestion originally made in Ref.~\onlinecite{jung2012ring}, assume the four overlapping strands are equivalent to that of a system of four linear polymers of the same length, each separately confined to a tube of $1/4$ the cross sectional area, i.e. tubes with an effective diameter of $D'=D/2$. Using the de~Gennes blob model for channel-confined polymers, the confinement free energy of the system is given by $F_I = 4F_{\rm bl} N' (D')^{-1/\nu}$, and so
\begin{eqnarray}
F_{\rm I} & = & F_{\rm bl}(1-f_{\rm large})2^{1/\nu}ND^{-1/\nu},
\end{eqnarray}
where $F_{\rm bl}$ is the free energy of a single blob, which is of order $k_{\rm B}T$. In an adjacent region, labelled II, the single loop of length $f_{\rm large}N$ can be viewed as two overlapping linear polymers of length $N''=f_{\rm large}N/2$, each of which effectively occupies a cylinder of diameter $D''=D/\sqrt{2}$ (i.e. half the cross-sectional area of the actual cylinder). The two regions are illustrated in Fig.~\ref{fig:appendix_model}(a). The free energy is thus $\Delta F_{\rm II} = 2F_{\rm bl}N''(D'')^{-1/\nu}$, and so
\begin{eqnarray}
F_{\rm II} & = &  F_{\rm bl} f_{\rm large}2^{1/\nu}ND^{-1/\nu}.
\end{eqnarray}
The total conformational free energy of two segregated polymers is thus $F_{\rm seg}=2(F_{\rm I}+F_{\rm II})$, which gives
\begin{eqnarray}
F_{\rm seg} = F_{\rm bl} ND^{-1/\nu} \left( f_{\rm large}2^{1+1/2\nu} + (1-f_{\rm large})2^{1+1/\nu}  \right) 
\nonumber \\
\end{eqnarray}

Now consider the case of two overlapping polymers. As in Ref.~\onlinecite{mitra2022topology}, assume that the small loops of each polymer overlap with each other and that the two large loops overlap with each other. As above, call these regions I and II, respectively. The four overlapping loops of region~I can be viewed as eight overlapping linear polymers, each of length $(1-f_{\rm large})N/2$ occupying effective tubes of diameter $D/2\sqrt{2}$. Likewise the two overlapping large loops in region~II can be viewed as four overlapping linear polymers, each of length $f_{\rm large}N$ confined to effective tubes of diameter $D/2$. The two overlap regions are illustrated in Fig.~\ref{fig:appendix_model}(b). Again using the de~Gennes blob model, it is easy to show that the total confinement free energy for this overlapping configuration is given by
\begin{eqnarray}
F_{\rm seg} = F_{\rm bl} ND^{-1/\nu} \left( f_{\rm large}2^{1+1/\nu} + (1-f_{\rm large})2^{1+3/2\nu}  \right). 
\nonumber \\
\end{eqnarray}
Consequently, the free energy barrier height, defined by $\Delta F = F_{\rm ov} - F_{\rm seg}$, is given by
\begin{eqnarray}
\Delta F/F_{\rm bl} & = & c(f_{\rm large}) ND^{-1/\nu},
\end{eqnarray}
where
\begin{eqnarray}
c(f_{\rm large}) & = & 2^{1+1/\nu} f_{\rm large} + 2^{1+3/2\nu} (1-f_{\rm large})
\nonumber \\
&& - 2^{1+1/2\nu} f_{\rm large} - 2^{1+1/\nu} (1-f_{\rm large}).
\label{eq:cfl}
\end{eqnarray}
Using $\nu=\frac{3}{5}$ and $f_{\rm large} = \frac{1}{2}$, this expression yields a result consistent with that of Ref.~\onlinecite{mitra2022topology}.

\bibliography{paper}
\bibliographystyle{apsrev4-1}

\end{document}